\documentclass[aps,prb,onecolumn,superscriptaddress,10pt,article,nofootinbib,longbibliography]{revtex4-2}
\usepackage{graphicx,amsfonts,amssymb,amsmath,hyperref,hypcap,enumerate,bm,color,siunitx}

\newcommand{\braket}[2]{\left\langle #1 | #2 \right\rangle}
\newcommand{\bra}[1]{\left\langle#1\right|}
\newcommand{\ket}[1]{\left|#1\right\rangle}

\usepackage{comment}

\newcommand{\comm}[2]{\left[#1,#2\right]}
\newcommand{\anticomm}[2]{\left\{#1,#2\right\}}

\newcommand{\half}{\frac{1}{2}}

\newcommand{\beq}{\begin{equation}}
\newcommand{\eneq}{\end{equation}}

\newcommand{\abs}[1]{\left|#1\right|}
\newcommand{\dm}[1]{\rho^{(#1)}}
\newcommand{\beqal}{\begin{equation} \begin{aligned}}
\newcommand{\eneqal}{\end{aligned} \end{equation}}
\newcommand{\bematrix}{\begin{pmatrix}}
\newcommand{\enmatrix}{\end{pmatrix}}

\def\qq{\bm{q}}
\def\kk{\bm{k}}

\def\aa{\bm{a}}

\def\bA{\bm{A}}

\def\qq{\bm{q}}

\def\aa{\bm{a}}

\def\nn{\bm{n}}
\def\xx{\bm{x}}

\def\EE{\bm{E}}
\def\BB{\bm{B}}
\def\MM{\bm{M}}
\def\JJ{\bm{J}}
\def\ha{\hat{a}}

\def\mA{{\mathcal{A}}}

\def\mO{{\mathcal{O}}}
\def\mM{{\mathcal{M}}}
\def\tomega{\tilde{\omega}}
\def\tj{\tilde{j}}
\def\tp{\tilde{p}}
\def\tE{\tilde{E}}
\def\tpsi{\tilde{\psi}}

\def\teta{\tilde{\eta}}
\def\tbeta{\tilde{\beta}}
\def\tn{\tilde{n}}
\def\tB{\tilde{B}}

\def\AAs{\AA$^2$}

\def\ea{{\it et al.}}

\begin{document}

\title{Magnetic-resonance-induced nonlinear current response in magnetic Weyl semimetals}

\author{Ruobing Mei}
\affiliation{Department of Physics, The Pennsylvania State University, University Park,  Pennsylvania 16802, USA}
\author{Chao-Xing Liu}
\email{Corresponding author: cxl56@psu.edu}
\affiliation{Department of Physics, The Pennsylvania State University, University Park,  Pennsylvania 16802, USA}

\date{\today}

\begin{abstract}
In this work, we propose a geometric nonlinear current response induced by magnetic resonance in magnetic Weyl semimetals. This phenomenon is in analog to the quantized circular photogalvanic effect \cite{de2017quantized} 
previously proposed for Weyl semimetal phases of chiral crystals. However, the nonlinear current response in our case can occur in magnetic Weyl semimetals where time-reversal symmetry, instead of inversion symmetry, is broken. The occurrence of this phenomenon relies on the special coupling between Weyl electrons and magnetic fluctuations induced by magnetic resonance. To further support our analytical solution, we perform numerical studies on a model Hamiltonian describing the Weyl semimetal phase in a topological insulator system with ferromagnetism. 
\end{abstract}

\maketitle



\section{Introduction}

Topological and geometric aspects of electronic Bloch wavefunctions in the Brillouin zone (BZ), including the Berry curvature \cite{berry1984quantal, xiao2010berry, liu2024quantum} and quantum metric \cite{resta2011insulating, torma2023essay, provost1980riemannian, liu2024quantum}, play essential roles in our understanding of physical response in crystalline materials. For example, the Hall conductivity of a 2D crystalline insulator can be related to the integral of the momentum-space Berry curvature for the Bloch wavefunctions over the entire BZ, which is identified as a topological index called the ``Chern number" in the formula first derived by Thouless, Kohmoto, Nightingale, and den Nijs \cite{thouless1982quantized}. The Berry curvature integral of the occupied states, which is not quantized, also provides an intrinsic mechanism for the anomalous Hall effect in a variety of ferromagnetic metals. More recently, physical phenomena induced by nonlinear response were also connected to band topology or geometry. For example, the Berry curvature dipole and quantum metric dipole are theoretically proposed to be the origin of the nonlinear Hall effect in non-magnetic or anti-ferromagnetic materials, which preserves either time-reversal symmetry ($\hat{T}$) or the combined $\hat{P}\hat{T}$-symmetry with inversion $\hat{P}$ \cite{sodemann2015quantum, du2021nonlinear, zhang2018berry, wang2021intrinsic, das2023intrinsic, kaplan2024unification}. This geometric origin of nonlinear Hall effect have been experimentally demonstrated in a variety of material compounds \cite{ma2019observation, kang2019nonlinear, shvetsov2019nonlinear, dzsaber2021giant, qin2021strain, tiwari2021giant, ma2022growth, huang2023intrinsic, huang2023giant, zhao2023berry, gao2023quantum, wang2023quantum}. Quantized topological nonlinear response due to chiral charges of Weyl nodes has also be theoretically proposed for the circular photogalvanic effect (CPGE) in Weyl semimetals \cite{de2017quantized, le2020ab, le2021topology}. 
Experimental observation of CPGE has been reported in chiral multifold semimetals RhSi and CoSi and its connection to topological chiral charges has been discussed \cite{rees2020helicity, ni2021giant, ni2020linear}.

\begin{figure} [htb]
    \centering
    \includegraphics[width=\textwidth]{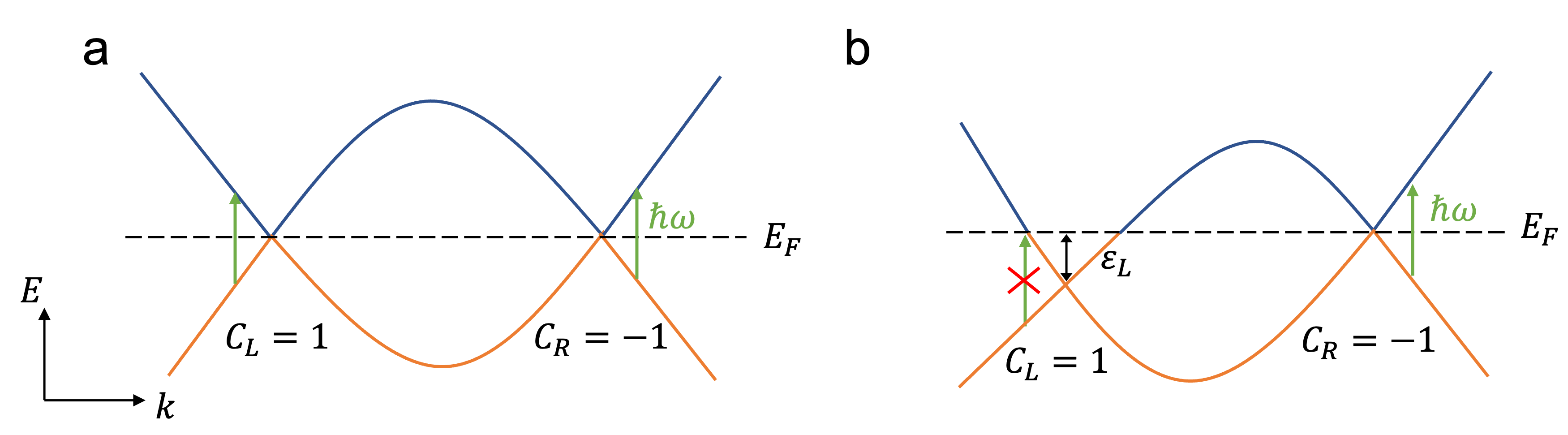}
    \caption{(a) Schematic of the energy dispersion for a two-band Weyl semimetal that preserves mirror symmetry and breaks time-reversal symmetry. The left and right Weyl nodes have opposite $C$ and sit at the same energy. The current response of topological CPGE will cancel between these two Weyl nodes due to the opposite chiral charges, but the magnetic-resonance-induced nonlinear current proposed in this work will remain. (b) Band structure of a chiral Weyl semimetal where the two Weyl nodes are at different energies. The topological CPGE can exist due to the Pauli blocking. }
    \label{fig:fig1}
\end{figure}

The CPGE is a second-order optical response, in which the current switches its direction when the circular polarized incident light changes its polarization \cite{de2017quantized, ji2019spatially, ni2021giant, ma2021topology, flicker2018chiral, rees2020helicity, ni2020linear, le2021topology}. In Ref. \cite{de2017quantized}, de Juan \ea \ proposed that the CPGE can be quantized in the Weyl semimetal phase of chiral crystals and this quantization originates from the topological chiral charge of Weyl nodes. Because the Weyl nodes must appear in pairs with opposite chiral charges in crystals, the CPGE contribution from the Weyl nodes with opposite chiral charges will cancel each other, generally leading to the vanishing CPGE (Fig. \ref{fig:fig1}a). To overcome this obstruction, it was theoretically proposed that when two Weyl nodes with opposite chiral charges sit at different energies and the Fermi energy is close to one Weyl node but far away from the other Weyl node (Fig. \ref{fig:fig1}b), optical transitions can only occur for the Weyl fermion close to the Fermi energy but not the other because of the Pauli blocking \cite{de2017quantized}. This scenario can occur in the Weyl semimetal phase of chiral crystals, in which the inversion and all the mirror symmetries are broken, so that no symmetry can connect the Weyl nodes with opposite chiral charges and thus they can appear at different energies. We note that time-reversal symmetry can only relate two Weyl nodes with the same chiral charge. 
However, experimental confirmation of the quantized CPGE is still lacking, and it was suggested that additional contributions from remote bands, rather than the two bands that form Weyl femrions, can have a significant contribution to the CPGE, which can be much larger than the quantized contribution in realistic materials \cite{zhang2018photogalvanic}.
Furthermore, it was also suggested that interactions and disorders can destroy the quantization of CPGE in chiral Weyl semimetals \cite{avdoshkin2020interactions, wu2024absence}.


In this work, we propose an alternative approach to overcome this obstruction in magnetic Weyl semimetals, in which time-reversal symmetry is broken but inversion symmetry is preserved so that two Weyl nodes with opposite chiral charges appear at the same energy. The CPGE contributions from different Weyl nodes will cancel each other so that we expect a vanishing CPGE for the nonlinear electromagnetic response in this case (Fig. \ref{fig:fig1}a). The key idea is to introduce the concept of ``pseudo-gauge field" via magnetic fluctuation or spin wave in magnetic Weyl semimetals \cite{yu2019magnetic,yu2021pseudo,ilan2020pseudo}. Here the pseudo-gauge field refers to the perturbation that behaves in a similar manner as a gauge field coupled to the Weyl fermions in the low energy sector of the system. Different from the usual electromagnetic field, this gauge coupling depends on the chirality of Weyl fermions, and has opposite signs when the Weyl fermion chirality reverses. In Sec. \ref{sec:pseudo}, we will study the nonlinear current response in magnetic Weyl semimetals due to the interplay between electromagnetic fields and the magnetic-resonance (MR) induced pseudo-gauge field, in analog to the CPGE purely from the electromagnetic fields. Strikingly, we find two Weyl nodes with opposite chiral charges contribute the same sign to the MR-induced nonlinear current, in sharp contrast to the opposite sign contributions of the normal CPGE. This difference originates from the chirality-dependent gauge coupling form of the pseudo-gauge field. In Sec. \ref{sec:LLG}, we will demonstrate that the magnetic fluctuation or spin wave, which can be driven via ferromagnetic resonance, serves as the pseudo-gauge field in magnetic Weyl semimetals.  In Sec. \ref{sec:microscopic}, we implemented numerical studies of the magnetic-resonance-induced nonlinear current response in a magnetic Weyl semimetal model that was previously adopted to describe magnetically doped topological insulators, in order to justify our analytical solutions.



\section{Model Hamiltonian} \label{sec:model-Hamiltonian}

We start from a four-band model that describe a topological insulator system with ferromagnetism in Ref. \cite{liu2013chiral}, which can capture the Weyl semimetal phase when magnetization term dominates over the nonmagnetic band gap. The model Hamiltonian reads
\beqal \label{eq:four-band}
 &H = H_0 +H_1, \\
 &H_0 = \mM(\kk) \tau_z + L_1 k_z \tau_y + L_2 (k_y \sigma_x - k_x \sigma_y) \tau_x + m \sigma_z, \\
 &H_1 = \bm{\nu} (t) \cdot \bm{\sigma} \tau_z,
\eneqal
where $\mM(\kk) = M_0 + M_1 k^2_z + M_2 (k^2_x + k^2_y)$, $M_{0,1,2}, L_{1,2}$ are material dependent parameters, and $m$ describes the out-of-plane magnetization and can induce the Weyl semimetal phase when $m>M_0$ by neglecting the quadratic terms in $\mM(\kk)$. 
Here we also include the $H_1$ term that represents the magnetic fluctuation and assume $\bm{\nu} (t)$ has temporal dependence. 
In Ref. \cite{liu2013chiral}, additional terms with the form $\bm{\mu} (t) \cdot \bm{\sigma}$ can exist for magnetic fluctuations. However, these terms will only slightly modify the form of the pseudo-gauge field discussed below and does not affect the final result. Thus, we drop these additional terms and only focus on the $\bm{\nu} (t)$ terms in this work. $H_1$ is treated as a perturbation below. Next we will show that when the model Hamiltonian is in the magnetic Weyl semimetal phase for $m>M_0$ and projected to the subspace of two low-energy Weyl fermions, the magnetic fluctuation in $H_1$ acts like a chiral gauge field which couples oppositely to the Weyl fermions with left and right chiralities \cite{liu2013chiral}. 

The eigen-energy of the Hamiltonian $H_0$ in Eq. (\ref{eq:four-band}) can be solved analytically as
\beq
\varepsilon_{\lambda\mu} = \lambda \sqrt{L_2^2 \left(k_x^2 + k_y^2 \right) + \left(\sqrt{\mM^2 + L_1^2 k_z^2} + \mu \abs{m} \right)^2},
\eneq
where $\lambda, \mu = \pm$, and we denote the eigenstates as $\ket{\lambda,\mu}$. When $k_x = k_y = 0$, the eigen-energies of the two middle bands are $\varepsilon_{-,-} = - \sqrt{\mM^2 + L_1^2 k_z^2} +m$ and $\varepsilon_{+,-} = \sqrt{\mM^2 + L_1^2 k_z^2} -m$, assuming $m > 0$. Thus, the gap between the two states $\ket{-,-}$ and $\ket{+,-}$ can be closed at $\mM^2 + L_1^2 k_z^2 = m^2$ if $m > \abs{\mM}$. By neglecting the quadratic term $M_1 k_z^2$ in $\mM (\kk)$, namely $\mM (\kk)=M_0$, we can  find two gap closing points at $\kk =(0, 0, \pm K_0)$ with $K_0 = \sqrt{m^2 - M_0^2}/L_1$.

At the gap closing point $\kk_0 = (0, 0, K_0)$, the Hamiltonian can be simplified as
\beq
H_0 (\kk_0) = M_0 \tau_z + L_1 K_0 \tau_y +m \sigma_z,
\eneq
and the eigenstates are solved as
\beqal
&\ket{-,+} = \frac{1}{\sqrt{N_-}} \begin{pmatrix}
    0 \\ 0 \\ M_0 - m \\ i L_1 K_0
\end{pmatrix},\quad \ket{-,-} = \frac{1}{\sqrt{N_-}} \begin{pmatrix}
    M_0 - m \\ i L_1 K_0 \\ 0 \\ 0
\end{pmatrix}, \\
&\ket{+,-} = \frac{1}{\sqrt{N_+}} \begin{pmatrix}
    0 \\ 0 \\ M_0 + m \\ i L_1 K_0
\end{pmatrix},\quad \ket{+,+} = \frac{1}{\sqrt{N_+}} \begin{pmatrix}
    M_0 + m \\ i L_1 K_0 \\ 0 \\ 0
\end{pmatrix},
\eneqal
where $N_\pm = (M_0 \pm m)^2 + L_1^2 K_0^2$ are the normalization factors. The effective Hamiltonian around $\kk_0$ up to the first order in $\delta \kk$ is written as
\beq
H' = L_1 \delta k_z \tau_y + L_2 (\delta k_y \sigma_x - \delta k_x \sigma_y) \tau_x + \bm{\nu} \cdot \bm{\sigma} \tau_z.
\eneq
We then project the effective Hamiltonian $H'$ into the subspace of the basis $\ket{-,-}$ and $\ket{+,-}$ by applying the second-order perturbation
\beq
H_{mn} = \bra{m} H' \ket{n} + \sum_{l \neq m,n} \half \bra{m} H' \ket{l} \bra{l} H' \ket{n} \times \left[\frac{1}{\varepsilon_m - \varepsilon_l} +\frac{1}{\varepsilon_n - \varepsilon_l} \right],
\eneq
which gives the effective Hamiltonian in the low-energy subspace as
\beqal
H_{eff}^+ = &-\frac{M_0}{m} \nu_z + \left(-L_2 \delta k_x + \frac{L_1 K_0}{m} \nu_x\right) \sigma_x + \left(-L_2 \delta k_y + \frac{L_1 K_0}{m} \nu_y\right) \sigma_y \\
&- \frac{1}{2m}\left(2L_1^2 K_0 \delta k_z + \nu_z^2 - \frac{M_0^2}{m^2} (\nu_x^2+\nu_y^2+\nu_z^2)\right) \sigma_z,
\eneqal
where we keep $\delta \kk$ up to the first order and $\bm{\nu}$ up to the second order. Similarly, we can perform the projection at the other gap closing point $(0,0,-K_0)$, and obtain the effective Hamiltonian
\beqal
H_{eff}^- = &-\frac{M_0}{m} \nu_z + \left(L_2 \delta k_x + \frac{L_1 K_0}{m} \nu_x\right) \sigma_x + \left(L_2 \delta k_y + \frac{L_1 K_0}{m} \nu_y\right) \sigma_y \\
&- \frac{1}{2m}\left(-2L_1^2 K_0 \delta k_z + \nu_z^2 - \frac{M_0^2}{m^2} (\nu_x^2+\nu_y^2+\nu_z^2)\right) \sigma_z.
\eneqal
Combining the two Hamiltonians, we can then write down the low-energy effective Hamiltonian for the two gap closing points or two Weyl points as
\beqal\label{eq:Heff1}
H_{eff} &= \left(L_2 \delta k_x \tau_z + \frac{L_1 K_0}{m} \nu_x\right) \sigma_x + \left(L_2 \delta k_y \tau_z+ \frac{L_1 K_0}{m} \nu_y\right) \sigma_y \\
&+ \frac{1}{2m}\left(2L_1^2 K_0 \delta k_z \tau_z - \nu_z^2 + \frac{M_0^2}{m^2} (\nu_x^2+\nu_y^2+\nu_z^2)\right) \sigma_z \\
& = \hbar v_f \bm{\sigma} \cdot \left(\delta \kk \, \tau_z + \frac{e}{\hbar} \aa \right),
\eneqal
where $\bm{\sigma}$ are Pauli matrices in the spin basis, $\tau_z$ is in the basis of two Weyl points, $\hbar v_f = L_2$, $\delta \kk = (\delta k_x, \delta k_y, \frac{L_1^2 K_0}{L_2 m}\delta k_z)$, and $\aa = (a_x, a_y, a_z)$ with $a_x = \frac{\hbar L_1 K_0}{e L_2 m} \nu_x$, $a_y = \frac{\hbar L_1 K_0}{e L_2 m} \nu_y$, and $a_z = \frac{\hbar}{2 e L_2 m}(- \nu_z^2 + \frac{M_0^2}{m^2} (\nu_x^2+\nu_y^2+\nu_z^2))$. We may introduce the electromagnetic field by the Peierls substitution $\delta \kk \rightarrow \delta \kk + \frac{e}{\hbar} \bA$ with the vector potential $\bA$. By comparing the coupling form of $\bA$ with $\aa$, we find the $\aa$ field, which results from the magnetic fluctuation $\bm{\nu}$, couples oppositely to the Weyl nodes at the two gap closing points in a gauge coupling form, and thus we conclude the $\aa$ field acts like a chiral gauge field \cite{liu2013chiral}. Below we will use the terminology, either chiral gauge field or pseudo-gauge field, to refer to the $\aa$ field. 

Due to the presence of magnetic ordering, the time-reversal symmetry is broken in magnetic Weyl semimetals, and the minimal model of Weyl semimetals with two Weyl nodes can be realized by the model Hamiltonian in Eq. (\ref{eq:four-band}), as shown by the numerical simulations of the energy bands for Eq. (\ref{eq:four-band}) in Fig. \ref{fig:fig2}a. 
On the other hand, if only inversion symmetry is broken, the total number of Weyl nodes in the system must be a multiple of four, as the time-reversal symmetry transforms a Weyl node at $\kk$ to another Weyl node at $-\kk$ with the same chirality and all the chiral charges for the whole Weyl semimetals must cancel \cite{armitage2018weyl}. 

\section{Magnetic-resonance-induced current response for Weyl fermions} \label{sec:pseudo}
We next consider the nonlinear response for the left-handed and right-handed Weyl fermions described by the Hamiltonian based on the effective Hamiltonian Eq. (\ref{eq:Heff1})
\begin{equation}
\begin{aligned}
 &H_L = \hbar v_f \bm{\sigma} \cdot \left(\kk + \frac{e}{\hbar} \bA  + \frac{e}{\hbar}\aa \right), \\
 &H_R = -\hbar v_f \bm{\sigma} \cdot \left(\kk + \frac{e}{\hbar} \bA  -\frac{e}{\hbar} \aa \right),
\end{aligned}
\end{equation}
where $\bm{\sigma}$ is Pauli matrices, $\bA$ is the electromagnetic field, and $\aa$ is the chiral gauge field minimally coupled to the left-handed and right-handed Weyl fermions \cite{liu2013chiral}. We rewrite the above Hamiltonian as
\begin{equation}
\begin{aligned}
 H_L = H_{L0} + \bA \cdot \JJ_L + \aa \cdot \JJ_L, \\
 H_R = H_{R0} + \bA \cdot \JJ_R - \aa \cdot \JJ_R, \\
\end{aligned}
\end{equation}
where $H_{L0} =-H_{R0} = \hbar v_f \bm{\sigma} \cdot \kk$ and $\JJ_L = - \JJ_R = ev_f \bm{\sigma}$. 
Let us first focus on the left-handed Weyl fermion $H_L$. Following the density matrix formalism in the velocity gauge described in Appendix \ref{sec:sm-velocity-gauge}, we rewrite the Hamiltonian into the second-quantization form as
\beq
 \hat{H}_L = \sum_n \int d\kk \varepsilon^L_{n\kk} \ha^{\dag}_{n\kk} \ha_{n\kk} +\sum_{nn'} \int d\kk \ha^{\dag}_{n\kk} \ha_{n'\kk} \left[ \bA(t)\cdot \JJ^L_{nn'\kk} + \aa_L (t) \cdot \JJ^L_{nn'\kk} \right],
\eneq
where $\varepsilon^L_{n\kk}$ is the eigenvalue of $H_{L0}$ and $\hat{a}_{n\kk}$ and $\hat{a}_{n\kk}^{\dag}$ are the annihilation and creation operators. The Bloch equation of the density matrix $\rho_{nm}(t)$, defined by $\rho_{nm,\kk} (t) = \langle \hat{a}_{m\kk}^{\dag}(t) \hat{a}_{n\kk} (t) \rangle$,
is then
\begin{equation}
\begin{aligned}
\frac{\partial \rho_{nm}(t)}{\partial t} = -\frac{i}{\hbar} \varepsilon^L_{nm} \rho_{nm}(t) - \frac{i}{\hbar} \sum_{n'} \left[\bA(t) \cdot \JJ^L_{n'n} \rho_{n'm}(t) - \bA(t) \cdot \JJ^L_{mn'} \rho_{nn'}(t) \right] \\
- \frac{i}{\hbar} \sum_{n'} \left[\aa_L(t) \cdot \JJ^L_{n'n} \rho_{n'm}(t) - \aa_L(t) \cdot \JJ^L_{mn'} \rho_{nn'}(t) \right] - \frac{\rho_{nm}(t)}{\tau},
\end{aligned}
\end{equation}
where $\tau$ is the relaxation time and $\varepsilon^L_{nm} = \varepsilon^L_n - \varepsilon^L_m$. We can expand the density matrix in order of the perturbation of dynamic fields $\bA$ and $\aa$. The zeroth order of the density matrix is given by $\dm{0}_{nm} = f_n \delta_{nm}$ where $f_n$ is the Fermi factor of band $n$. 
At the first order of $\bA$ and $\aa$, we obtain
\beq
\dm{1}_{nm} (\omega) =f_{nm} \frac{ \left(A^b (\omega) + a^b_L (\omega) \right) (J_L)^b_{mn}}{\hbar (\tomega_{1,nm} - \omega)},
\eneq
where we define $f_{nm} = f_n - f_m$, $A^b (t) = A^b (\omega)  e^{-i\omega t}$, $a_L^b (t)  = a_L^b (\omega) e^{-i\omega t}$, and $\tomega_{1,nm} =\varepsilon^L_{nm}/\hbar - i/\tau $. 
Here the Einstein summation rule has been assumed. At the second order of $\bA$ and $\aa$, the diagonal component of density matrix can be derived as 
\beqal
\dm{2}_{nn} (\omega_{\Sigma}) = \frac{1}{\hbar \omega_{\Sigma}}\sum_m \left(A^b_{\beta} (a_L)^c_{\gamma} + (a_L)^b_{\beta} A^c_{\gamma} \right) \left[ \frac{f_{mn}(J_L)^b_{mn} (J_L)^c_{nm} }{\hbar (\tomega_{1,mn} - \omega_{\beta})} - \frac{f_{nm}(J_L)^b_{nm} (J_L)^c_{mn} }{\hbar (\tomega_{1,nm} - \omega_{\beta})} \right],
\eneqal
where $\omega_{\Sigma} = \omega_{\beta} + \omega_{\gamma}$, $A^b_{\beta} = A^b(\omega_{\beta})$, and $(a_L)^b_{\beta} = (a_L)^b (\omega_{\beta})$. 
Here we only consider the diagonal term in the second order of density matrix $\rho^{(2)}$. This is because we only consider two bands that form the Weyl points here, and the off-diagonal component of $\rho^{(2)}$ is zero. It should be noted that if the third bands are involved, the off-diagonal component of $\rho^{(2)}$ can exist and a three-band contribution can be important for the CPGE \cite{zhang2018photogalvanic}. 
Moreover, we only keep the terms involving both $\bA$ and $\aa_L$, as the other terms quadratic in $\bA$ or $\aa_L$ vanish between the left-handed and right-handed Weyl fermions due to their opposite chiral charges (see Appendix \ref{sec:sm-length-gauge}). The corresponding intraband current is derived as
\beqal
 \tj_L^a (t) &= e \sum_n \int_k  v^a_{nn} \dm{2}_{nn} (\omega_{\Sigma}) e^{-i\omega_{\Sigma} t} \\
 &=\frac{e}{\hbar^2 \omega_{\Sigma}}\sum_{nm} \int_k \frac{\Delta^a_{nm} f_{mn}(J_L)^b_{mn} (J_L)^c_{nm} }{\tomega_{1,mn} - \omega_{\beta}} \left(A^b_{\beta} (a_L)^c_{\gamma} + (a_L)^b_{\beta} A^c_{\gamma} \right)e^{-i\omega_{\Sigma} t},
\eneqal
where $v^a_{nm} = \bra{n} \frac{\partial H}{\partial k_a} \ket{m}$ is the matrix element of the velocity operator and $\Delta^a_{nm} = v^a_{nn} - v^a_{mm}$ is the velocity shift. We then use $A^b (\omega) = E^b(\omega)/i\omega$, $a_L^b (\omega)= \tE^b (\omega)/i\omega $ with the pseudo-electric field $\tE$, and $(J_L)^b_{nm} = \frac{ie}{\hbar}\varepsilon^L_{nm} \mA^b_{nm}$ with the Berry connection $\bm{\mA}_{nm} = \bra{n} i\partial_{\kk} \ket{m}$. We symmetrize all the indices, and take the time derivative of the imaginary part (see Appendix \ref{sec:sm-length-gauge}) to obtain
\beq
 \frac{d \tj^a_L (t)}{dt} = \teta^L_{abc} (\omega_{\Sigma};\omega_{\beta}, \omega_{\gamma}) E^b_{\beta} \tE^c_{\gamma} e^{-i\omega_{\Sigma} t} + \teta^L_{abc} (\omega_{\Sigma
 };\omega_{\gamma}, \omega_{\beta}) \tE^b_{\beta} E^c_{\gamma} e^{-i\omega_{\Sigma} t},
\eneq
where
\beqal
 \teta^L_{abc} (\omega_{\Sigma};\omega_{\beta}, \omega_{\gamma}) &= -\frac{i \pi e^3}{8\hbar^2}  \sum_{nm} \int_k \Delta^a_{nm} f_{nm} (\Omega_L)^{bc}_{mn} \frac{\varepsilon^2_{nm}}{\hbar^2 \omega_{\beta} \omega_{\gamma}} \\
 & \times \left[\delta(\omega_{mn} - \omega_{\beta}) - \delta(\omega_{mn} + \omega_{\beta}) - \delta(\omega_{mn} - \omega_{\gamma}) + \delta(\omega_{mn} + \omega_{\gamma}) \right],
\eneqal
with the Berry curvature of the left-handed Weyl fermion $(\Omega_L)^{bc}_{mn} = i(\mA^b_{mn}\mA^c_{nm} - \mA^c_{mn}\mA^b_{nm})$. For $\omega_{\beta} = -\omega_{\gamma} = \omega$ and $\omega_{\Sigma} = 0$, we have
\beq
 \teta^L_{abc}(0; \omega,-\omega) = -\frac{i \pi e^3}{2\hbar^2}  \sum_{nm} \int_k \Delta^a_{nm} f_{nm} (\Omega_L)^{bc}_{mn} \delta (\omega_{mn} - \omega).
\eneq
We rewrite the above equations as (see Appendix \ref{sec:sm-length-gauge})
\beq \label{eq:current2}
 \tilde{\bm{j}}_L = \tau \tbeta_L (\omega) \left[\EE(\omega) \times \tilde{\EE}^* (\omega) +\tilde{\EE}(\omega) \times \EE^* (\omega)\right],
\eneq
where $\tilde{\EE} (\omega) = i\omega \aa$ can be regarded as the pseudo-electric field, and the MR-induced current response trace for the left-handed Weyl fermion is
\beqal
 \tbeta_L (\omega) &= i\frac{e^3}{2h^2} \oint d\bm{S} \cdot \bm{\Omega}_L = i\beta_0 C_L,
\eneqal
where $\bm{S}$ is a closed surface in the momentum space, $\beta_0 = \pi e^3/h^2$, and $C_L = \frac{1}{2\pi} \oint d\bm{S} \cdot \bm{\Omega}_L = 1$ is nothing but the Chern number (equivalently the chiral charge of the Weyl node). 
For the right-handed Weyl fermion, we 
repeat the above derivation to obtain
\beq
\tilde{\bm{j}}_R = - \tau \tbeta_R (\omega) \left[\EE(\omega) \times \tilde{\EE}^* (\omega) +\tilde{\EE}(\omega) \times \EE^* (\omega)\right],
\eneq
with
\beq
\tbeta_R (\omega) = i\beta_0 C_R = -i\beta_0 C_L.
\eneq
Thus, the total nonlinear response current is \beqal \label{eq:effective-CPGE}
\tilde{\bm{j}} &= \tilde{\bm{j}}_L + \tilde{\bm{j}}_R = 2i \tau\beta_0 C_L \left[\EE(\omega) \times \tilde{\EE}^* (\omega) +\tilde{\EE}(\omega) \times \EE^* (\omega)\right].
\eneqal
The current equation (\ref{eq:effective-CPGE}) is the main result of this work, from which we demonstrate that a non-zero total {\it dc}-current can be induced by the interplay between an electromagnetic field and a chiral gauge field at the nonlinear order for two Weyl nodes with opposite chiral charges, in contrast to the regular CPGE as discussed in Appendix \ref{sec:sm-length-gauge} (also see Ref. \cite{de2017quantized}). From Eq. (\ref{eq:Heff1}), we can see that the chiral gauge field $\aa$ depends on the time-dependent field $\bm{\nu} (t)$ that describes magnetic fluctuations and thus we study the magnetic dynamics in magnetic topological insulators below.

\section{Landau-Lifshitz-Gilbert equation for ferromagnetic resonance} \label{sec:LLG}

To see how the magnetic fluctuation can induce a pseudo-electric field, we solve the Landau-Lifshitz-Gilbert (LLG) equation for ferromagnetic resonance in this section. Similar to Ref. \cite{yu2019magnetic}, the LLG equation reads
\beqal
 \frac{d \MM}{dt} = -\gamma_0 \MM \times \left( \BB_{\text{eff}} - \eta \frac{d \MM}{dt} \right),
\eneqal
where $\MM = (M_x, M_y, M_z)$ is the magnetization. The in-plane magnetization acts as a pseudo-gauge field in terms of $\bm{\nu} = \frac{c }{e v_f} g_M(-M_y, M_x, 0)$, with the exchange coupling coefficient $g_M$ \cite{yu2019magnetic}. Therefore, the pseudo-electric field induced by the ferromagnetic resonance is 
\beq \label{eq:pseudo_E_field}
\tilde{\EE} = -\frac{1}{c} \frac{\partial \bm{\nu}}{\partial t} = \frac{g_M}{e v_f} (\frac{d M_y}{dt}, -\frac{dM_x}{dt}, 0).
\eneq
The LLG equation can be rewritten as
\beqal
 \frac{d \MM}{dt} &= -\gamma_0 \MM \times \left[\BB_{\text{eff}} - \eta \left( -\gamma_0 \MM \times \left( \BB_{\text{eff}} - \eta \frac{d \MM}{dt} \right) \right) \right] \\
 & \approx -\gamma_0 \MM \times \BB_{\text{eff}} - \eta \gamma^2_0 \MM \times (\MM \times \BB_{\text{eff}}),
\eneqal
where $\BB_{\text{eff}} = \BB - \MM_N - \frac{KM_z}{M^2_s} \hat{e}_z$, $\BB$ is the applied magnetic field, $M_s = \abs{\MM}$, $K$ is the anisotropy coefficient, $\MM_N = (N_x M_x, N_y M_y, N_z M_z)$ is the demagnetizing field with $N_x, N_y = 0$, $\gamma_0 = \frac{ge}{2m_e c}$ is the magneto-mechanical ratio with Land\'e factor $g$, and $\eta$ is Gilbert damping coefficient \cite{yu2019magnetic,kittel1976introduction,aharoni2000introduction}.
Let us define the dimensionless damping constant $\alpha = \gamma_0 \eta M_s$, $\gamma = \frac{\gamma_0}{1 + \alpha^2}$, and the LLG equation is written as
\beq
\frac{d\MM}{dt} = -\gamma \MM \times \BB_{\text{eff}} - \frac{\gamma \alpha}{M_s} \MM \times (\MM \times \BB_{\text{eff}}).
\eneq
Since $dM_s/dt = 0$, only the direction $\nn = \MM/M_s$ is time-dependent with
\beq
\frac{d\nn}{dt} = -\gamma \nn \times \BB_{\text{eff}} - \gamma \alpha \nn \times (\nn \times \BB_{\text{eff}}).
\eneq
We assume the in-plane magnetic field and magnetization are very small, i.e. $\abs{n_x} \sim \abs{n_y} \sim \abs{B_x/M_s} \sim \abs{B_y/M_s} \ll 1 $, and at the first order of these quantities, we have
\begin{equation} \label{eq:LLG-1}
\begin{pmatrix}
   \partial_t + \alpha \tomega & \tomega n_z \\
   -\tomega n_z & \partial_t + \alpha \tomega
\end{pmatrix} 
\begin{pmatrix}
    n_x \\ n_y
\end{pmatrix} = 
\begin{pmatrix}
    \gamma \alpha & \gamma n_z \\
    -\gamma n_z & \gamma \alpha
\end{pmatrix}
\begin{pmatrix}
    B_x \\ B_y
\end{pmatrix},
\end{equation}
where $\tomega = \gamma n_z B_{\text{eff},z}$ and $n_z = M_z/ M_s = \text{sgn}(M_z)$. We then perform the Laplace transformation as
\beqal
\tn_i (s) = \int_0^{\infty} dt e^{-st} n_i (t), \\
\tB_i (s) = \int_0^{\infty} dt e^{-st} B_i (t). \\
\eneqal
Eq. (\ref{eq:LLG-1}) then becomes
\beq \label{eq:LLG-2}
G_0 (s) \begin{pmatrix}
    \tn_x (s) \\ tn_y (s) 
\end{pmatrix} - \begin{pmatrix}
    n_x (0) \\ n_y (0)
\end{pmatrix} = G_1 \begin{pmatrix}
    \tB_x (s) \\ \tB_y (s)
\end{pmatrix},
\eneq
where
\beq
 G_0 (s) = \begin{pmatrix}
    s+\alpha \tomega & \tomega n_z \\
    -\tomega n_z & s+\alpha \tomega 
\end{pmatrix}, \
G_1 = \begin{pmatrix}
    \gamma \alpha & \gamma n_z \\
    -\gamma n_z & \gamma \alpha
\end{pmatrix}.
\eneq
We consider a general in-plane magnetic field $\BB (t) = B_0 e^{i\omega t} (a,b,0)$ where $a,b$ are two complex numbers and describe the polarization of magnetic field components of a microwave. 
The transformed field reads
\beqal
\tB_x (s) = \frac{B_0}{2} \left(\frac{a}{s - i\omega} + \frac{a^*}{s + i\omega} \right), \\
\tB_y (s) = \frac{B_0}{2} \left(\frac{b}{s - i\omega} + \frac{b^*}{s + i\omega} \right).
\eneqal
We then perform the inverse Laplace transformation on Eq. (\ref{eq:LLG-2}) and obtain
\beqal
 \begin{pmatrix}
     n_x \\ n_y 
 \end{pmatrix} & = \frac{B_0}{2} \left[ a e^{i\omega t} G_0^{-1} (i \omega) + a^*  e^{-i\omega t} G_0^{-1} (-i\omega)  \right] G_1 \bematrix 1 \\ 0 \enmatrix \\
 & + \frac{B_0}{2} \left[ b e^{i\omega t} G_0^{-1} (i \omega) + b^*  e^{-i\omega t} G_0^{-1} (-i\omega)  \right] G_1 \bematrix 0 \\ 1 \enmatrix, 
\eneqal
where
\beq
G_0^{-1} (s) = \frac{1}{(s + \alpha \tomega)^2 + \tomega^2} \bematrix s+ \alpha \tomega & -\tomega n_z \\ \tomega n_z & s + \alpha \tomega \enmatrix.
\eneq
The solutions to the above equation are
\beqal
&n_x (t) = \frac{B_0 \gamma}{2D(\omega)} \left[A_1 (\omega)\cos(\omega t) + A_2 (\omega) \sin (\omega t) \right] \\
A_1(\omega) &= ((\alpha^2 +1) \tomega^2 - \omega^2) \left[(a + a^*) (\alpha^2 +1) \tomega + i\omega ((a-a^*) \alpha + (b - b^*)n_z)\right] \\
& - 2i\alpha \omega \tomega \left[ (a-a^*)(\alpha^2 +1) \tomega + i\omega((a+a^*)\alpha + (b+b^*) n_z) \right] \\
A_2(\omega) &= ((\alpha^2 +1) \tomega^2 - \omega^2) \left[i(a - a^*) (\alpha^2 +1) \tomega - \omega ((a+a^*) \alpha + (b + b^*)n_z)\right] \\
& - 2i\alpha \omega \tomega \left[ i(a+a^*)(\alpha^2 +1) \tomega + \omega((a^* -a)\alpha + (b^*-b) n_z) \right] ,
\eneqal
and
\beqal
&n_y (t) = \frac{B_0 \gamma}{2D(\omega)} \left[A_3 (\omega)\cos(\omega t) + A_4 (\omega) \sin (\omega t) \right] \\
A_3(\omega) &= ((\alpha^2 +1) \tomega^2 - \omega^2) \left[(b + b^*) (\alpha^2 +1) \tomega + i\omega ((b-b^*) \alpha + (a^* - a)n_z)\right] \\
& - 2i\alpha \omega \tomega \left[ (b-b^*)(\alpha^2 +1) \tomega + i\omega((b+b^*)\alpha - (a+a^*) n_z) \right] \\
A_4(\omega) &= ((\alpha^2 +1) \tomega^2 - \omega^2) \left[i(b - b^*) (\alpha^2 +1) \tomega - \omega ((b+b^*) \alpha - (a + a^*)n_z)\right] \\
& - 2i\alpha \omega \tomega \left[ i(b+b^*)(\alpha^2 +1) \tomega + \omega((b^* -b)\alpha + (a-a^*) n_z) \right] ,
\eneqal
where
\beq
D(\omega) = (\alpha^2 +1)^2 \tomega^4 + \omega^4 + 2(\alpha^2 -1) \tomega^2 \omega^2.
\eneq
We find the resonance frequency satisfying $D(\omega_r) = 0$ to be
\beq
\omega_r = \sqrt{\frac{\sqrt{(\alpha^2 +1)^2(4\alpha^2 +1)} - (\alpha^2 +1)^2}{\alpha^2}} \abs{\tomega}.
\eneq
We keep the leading order $\mO(\alpha^{-1})$ to get
\beqal
&n_x (t) = \frac{B_0 \gamma_0}{2\alpha \omega_0} \left[( a_{\text{Im}} +  b_{\text{Re}} n_z) \cos (\omega_0 t) + (a_{\text{Re}} - b_{\text{Im}} n_z) \sin (\omega_0 t) \right], \\
&n_y (t) = \frac{B_0 \gamma_0}{2\alpha \omega_0} \left[( b_{\text{Im}} - a_{\text{Re}} n_z) \cos (\omega_0 t) + (b_{\text{Re}} + a_{\text{Im}} n_z) \sin (\omega_0 t) \right],
\eneqal
where $\omega_r \approx \omega_0 = \abs{\tomega}$, and Re and Im denote the real and imaginary parts of $a$ and $b$, respectively. Therefore, from Eq. (\ref{eq:pseudo_E_field}), 
the pseudo-electric field reads 
\beqal
 &\tE_x = \frac{g_M}{e v_f} \frac{d M_y}{d t}= \frac{g_M M_s}{e v_f} \dot{n}_y = \tE_0  (b - i a n_z) e^{i\omega_0 t}, \\
 &\tE_y = -\frac{g_M}{e v_f} \frac{d M_x}{d t}= - \frac{g_M M_s}{e v_f} \dot{n}_x = -\tE_0  (a + i b n_z) e^{i\omega_0 t},
\eneqal
where $\tE_0 = g_M B_0 \gamma_0 M_s/(2 e v_f \alpha)$. Thus, an in-plane magnetic field $\BB (t) = B_0 e^{i\omega_0 t} (a,b,0)$ can induce a pseudo-electric field
\beq\label{eq:ChiralElectricField}
\tilde{\EE} (t) = \tE_0 e^{i\omega_0 t} (b- ia n_z, -a - i b n_z, 0),
\eneq
where $n_z = \text{sgn}(M_z)$.
Combining Eq. (\ref{eq:ChiralElectricField}) with Eq. (\ref{eq:effective-CPGE}) for the nonlinear current response and considering an incident light with $\EE(t) = E_0 e^{i\omega_0 t} (c, d, 0)$, we find 
\beqal \label{eq:current}
\tilde{\bm{j}} &= 2i \tau\beta_0 C_L E_0 \tE_0 \left[ (-a^* + ib^*n_z) c - (b^* + ia^*n_z) d - (-a - ibn_z) c^* + (b - ian_z) d^* \right] \hat{e}_z  \\
&= - 4 \tau\beta_0 C_L E_0 \tE_0 \text{Im} \left[ (-a^* + ib^*n_z) c - (b^* + ia^*n_z) d \right] \hat{e}_z.  
\eneqal
Because the MR-induced nonlinear current response is generated by the combination of an electromagnetic field and a pseudo-electric field, the incident light can be linearly polarized, instead of circularly polarized for the regular CPGE. For example, we can choose an incident light with $\EE (t) = E_0 e^{i\omega_0 t} (1,0,0)$ and a magnetic field $\BB (t) = B_0 e^{i\omega_0 t}(0,1,0)$ to induce a pseudo-electric field $\tilde{\EE} (t) = \tE_0 e^{i\omega_0 t} (1,-i,0)$. Thus, using Eq. (\ref{eq:current}), the MR-induced nonlinear current is $\tilde{\bm{j}} = -4\tau \beta_0 C_L E_0 \tE_0 \hat{e}_z$ assuming $n_z = 1$. It should be noted that although the topological invariant $C_L$ appears in this current response, the pseudo-electric field $\tE_0$ depends on material parameters, not just the fundamental constants, and thus we regard this nonlinear current response to be geometric rather than topological. 


\section{Magnetic-resonance-induced nonlinear current response in a microscopic model} \label{sec:microscopic}


The above analytical derivation reveals the nonlinear current response induced by MR, in analog to the regular CPGE. In this section, we will further study the nonlinear current response directly from numerical calculations of the Hamiltonian Eq. (\ref{eq:four-band}). We emphasize that the purpose of this section is to verify the analytical solutions of the nonlinear current response in the low frequency regime for a more realistic Weyl semimetal model in certain limit. However, the parameters chosen for the calculations of this model are not realistic, so our calculations {\it cannot} be directly applied to the realistic material systems. A comment of the additional contribution to the nonlinear current response from the remote bands will be discussed at the end of this section.

As shown in Appendix \ref{sec:sm-pseudo-CPGE}, the MR-induced nonlinear current in the four-band model Eq. (\ref{eq:four-band}) can be derived as
\beq \label{eq:four-band-CPGE}
 \tj_a = \tau \teta_{abc} (0;\omega,-\omega) E_b (\omega) \tE_c^* (\omega) + \tau \teta_{acb} (0;-\omega, \omega) \tE_b (\omega) E_c^* (\omega),
\eneq
with
\beqal \label{eq:eta_2}
 \teta_{abc} (0;\omega,-\omega) 
 = i \beta_0 
 \sum_{nm} \int \frac{d^3 k}{2\pi} (\partial_a \varepsilon_{nm}) f_{nm} \frac{\mA^b_{nm} \Gamma^c_{mn}}{\varepsilon_{nm}} \delta (\varepsilon_{mn} - \hbar \omega),
\eneqal
where $\beta_0 = \pi e^3/h^2$, $\bm{\mA}_{nm}$ is the non-Abelian Berry connection of the unperturbed Hamiltonian $H_0$, $\bm{\Gamma} = \bm{\sigma} \tau_z$, and $\tilde{\EE}(\omega) = i\hbar \omega \bm{\nu}/e$  is the pseudo-electric field induced by the magnetic fluctuation $\bm{\nu}$, 
which comes from the ferromagnetic resonance by solving the LLG equation in Sec. \ref{sec:LLG}. 

At low energy, the four-band Hamiltonian can be projected to the effective Hamiltonian describing two Weyl points, and the magnetic fluctuation $\bm{\nu}$ appears in the form of the chiral gauge field $\aa$ (see Sec. \ref{sec:model-Hamiltonian}). In this approximation, we can rewrite $\tilde{\EE} (\omega)$ as $i\omega \aa$ following the definition of Eq. (\ref{eq:current2}) 
and Eq. (\ref{eq:eta_2}) becomes
\beqal
 \teta_{abc} (0;\omega,-\omega) 
 = i \beta_0 \frac{\hbar \nu_c}{e a_c} 
 \sum_{nm} \int \frac{d^3 k}{2\pi} (\partial_a \varepsilon_{nm}) f_{nm} \frac{\mA^b_{nm} \Gamma^c_{mn}}{\varepsilon_{nm}} \delta (\varepsilon_{mn} - \hbar \omega),
\eneqal
where the pseudo-gauge field $\aa$ in terms of $\bm{\nu}$ is defined in Sec. \ref{sec:model-Hamiltonian}. Furthermore, Eq. (\ref{eq:four-band-CPGE}) can be recovered to Eq. (\ref{eq:effective-CPGE}) derived from the effective model for the two Weyl fermions by performing projection to the subspace of the two low-energy bands. This indicates that at the low-energy range, the MR-induced nonlinear current trace of this microscopic model should have a quantized value of $2i\beta_0$.

As the magnetization in our model is in the $z$ direction, only an in-plane pseudo-electric field can be induced by the ferromagnetic resonance \cite{yu2019magnetic}. We assume the incident light propagates in the $z$ direction and has in-plane electromagnetic fields, so the indices $b,c$ can only be $x$ or $y$, and $b\neq c$ as the electric fields are along perpendicular directions in CPGE. Furthermore, the four-band model has mirror $M_z$ symmetry which imposes constraints $\teta_{xxy} = \teta_{xyx} = \teta_{yxy} = \teta_{yyx} = 0$, and the in-plane rotational symmetry which imposes $\teta_{zxy} = -\teta_{zyx}$. Therefore, in the following calculations, we focus on the component $\teta_{zxy}$ given by
\beq
 \teta_{zxy} = i \beta_0 \frac{L_2 m}{\sqrt{m^2 - M_0^2}} \sum_{nm} \int \frac{d^3 k}{2\pi} (\partial_z \varepsilon_{nm}) f_{nm} \frac{\mA^x_{nm} \Gamma^y_{mn}}{\varepsilon_{nm}} \delta (\varepsilon_{mn} - \hbar \omega),
\eneq
where we have used $a_y = \frac{\hbar L_1 K_0}{e L_2 m} \nu_y$
and $K_0 = \sqrt{m^2 - M_0^2}/L_1$ from Sec. \ref{sec:model-Hamiltonian} to derive this result. 

\begin{figure} [htb]
    \centering
    \includegraphics[width=0.8\textwidth]{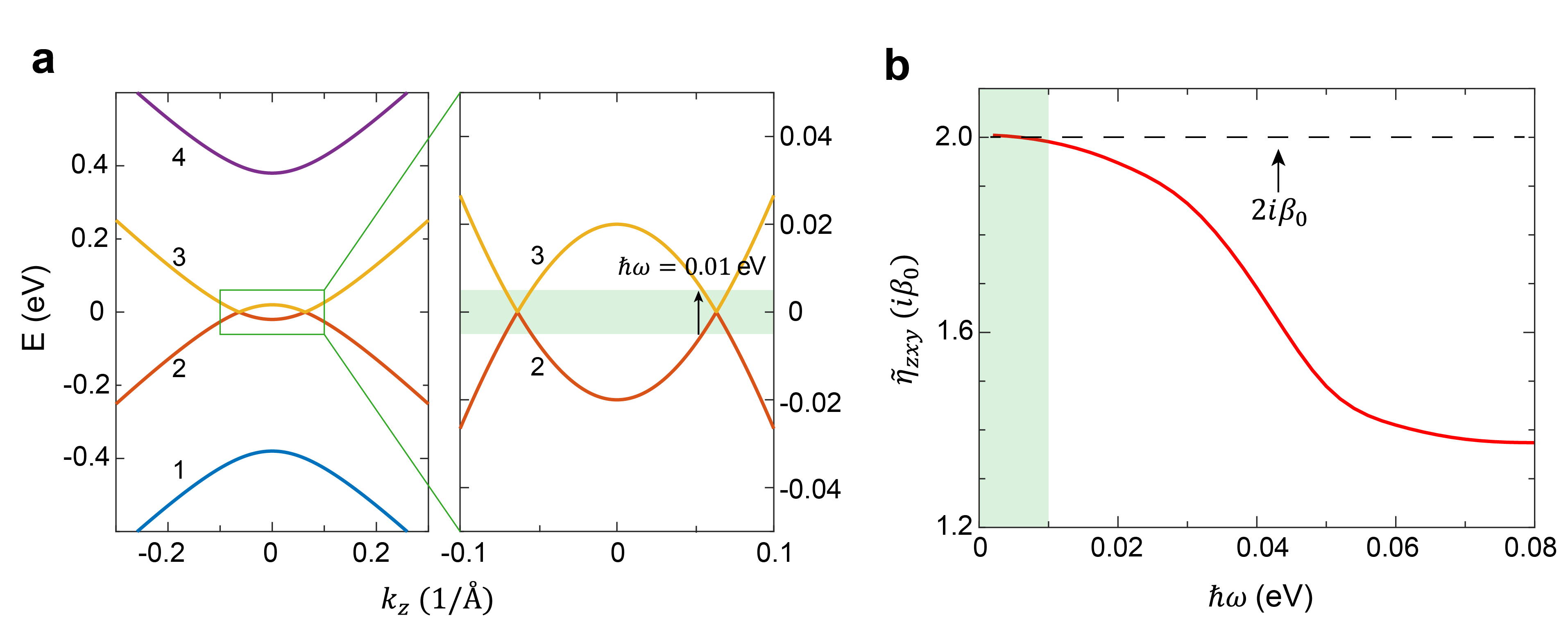}
    \caption{(a) 
    Energy dispersion of the four-band model along $k_z$ at $k_x = k_y = 0$ and the enlarged area around the two Weyl nodes. (b) The MR-induced nonlinear current component $\teta_{zxy}$ as a function of $\hbar \omega$. The quantized value is $2i \beta_0$, which can be achieved for $\hbar \omega < 0.01$ eV. The parameters are chosen as $M_0 = 0.18$ eV, $M_1 = 0.342$ eV\AAs, $M_2 = 18.25$ eV\AAs, $L_1 = 1.33$ eV\AA, $L_2 = 2.82$ eV\AA, $m = 0.2$ eV (parameters from Ref. \cite{liu2013chiral}). }
    \label{fig:fig2}
\end{figure}

Fig. \ref{fig:fig2}a shows the energy dispersion of the four-band model along $k_z$ at $k_x = k_y = 0$, in which four bands are labelled by the index $n=1, ..., 4$ from low to high energies. Two Weyl nodes with opposite charges at opposite $k_z$ are formed by two bands $n=2$ and $3$ around the energy $\varepsilon=0$. 
We numerically calculate $\teta_{zxy}$ as a function of $\hbar \omega$, as shown in Fig. \ref{fig:fig2}b. 
When $\hbar \omega$ is smaller than 0.01 eV (green shaded area), $\teta_{zxy}$ is very close to the quantized value $2i\beta_0$, suggesting that the microscopic model can be simplified as an effective model of two Weyl fermions at the low energy range. 
At around $\hbar \omega \sim 0.01$ eV, $\teta_{zxy}$ starts to decrease as the effective model of two Weyl fermions is no longer applicable, and thus $\teta_{zxy}$ deviates from quantization.
Another reason for the rapid decrease in $\teta_{zxy}$ is that the transitions around the Weyl nodes becomes only partially activated as $\hbar \omega$ exceeds the energy difference between the middle two bands at $k_z = 0$ ($\Delta \varepsilon \sim 0.04$ eV). Eventually, $\teta_{zxy}$ becomes a constant for $\hbar \omega > 0.06$ eV. Therefore, our numerical result confirms that the MR-induced nonlinear current trace between two bands of Weyl nodes can be quantized at the low energy range, where the microscopic model recovers to the effective model of two Weyl fermions, as shown in Sec. \ref{sec:model-Hamiltonian}.

In our calculation, we only considered two-band transitions ($2\to 3$, where $2, 3$ label two bands around $\varepsilon=0$ for two Weyl nodes) and have not included the three-band virtual transitions ($2 \to l \to 3$ where $l=1, 4$ labels the third bands for the three-band contributions to the nonlinear current response). It is known that the three-band contribution can play an important role for the CPGE when the third band $l$ is close to the two bands that form Weyl nodes \cite{zhang2018photogalvanic}, and thus destroys the quantization of CPGE. Similar physics can also occur for the MR-induced nonlinear current response here. The ratio between the three-band and two-band contributions can contain the energy separation ratio $\sim \frac{\varepsilon_n - \varepsilon_m}{\varepsilon_n - \varepsilon_l}$ \cite{zhang2018photogalvanic}, where $n,m = 2, 3$ and $l=1, 4$. This ratio decays by increasing the energy separation between the third band and two Weyl bands. For the chosen model parameters, a typical energy separation between two bands of Weyl nodes is around $0.01 \sim 0.02 eV$, below which the linear dispersion of Weyl fermions is valid. This energy scale is one order smaller than the energy separation between the third band and these two Weyl bands, which is around $0.2 eV$. 
As long as the other bands are far away enough from the two Weyl bands, we expect the contribution from the three-band transition provides a small correction. 


\section{Conclusion and discussion}

In conclusion, we demonstrate that a nonlinear current response with an electric field and a MR-induced pseudo-electric field can exist in magnetic Weyl semimetals with two Weyl nodes with opposite charges, in contrast to the regular CPGE. The pseudo-electric field can be induced by the magnetic fluctuation from ferromagnetic resonance, which acts like a chiral gauge field that couples oppositely to the lefthanded and righthanded Weyl nodes. Therefore, the realization of MR-induced nonlinear current does not require a chiral Weyl semimetal in which the two Weyl nodes are located at different energies. By combining this MR-induced nonlinear current response with the normal CPGE measurement, we anticipate a potential approach to extract the contribution from the quantized nonlinear response in realistic materials, which has not been achieved in the current experiments of normal CPGE measurements and will be left for the future work. 


\section{Acknowledgement}
We thank Binghai Yan and Jiabin Yu for helpful discussion. R. M and C.X.L acknowledge the support from the NSF through The Pennsylvania State University Materials Research Science and Engineering Center [DMR-2011839]. 

\appendix

\renewcommand\thefigure{\thesection.\arabic{figure}} 
\setcounter{figure}{0}

\section{Derivation of CPGE in length gauge} \label{sec:sm-length-gauge}

In this section, we derive the expression for CPGE current in the length gauge, in which the dipole interaction is treated as \cite{aversa1995nonlinear,ventura2017gauge,sipe2000second}
\begin{equation}
    \hat{H}_E(t) = \int d\xx \tilde{\psi}^{\dag} (\xx) \left[ H_0 + e\xx \cdot \EE (t) \right] \tilde{\psi} (\xx),
\end{equation}
where $\xx$ is the position operator and we define $H_1(t) = e\xx \cdot \EE (t)$. $H_0$ is the unperturbed single-particle Hamiltonian with the eigen-equation
\beq
 H_0 (\kk) \psi_{n\kk} (\xx) = \varepsilon_{n\kk} \psi_{n\kk}(\xx),
\eneq
where $\varepsilon_{n\kk}$ is the eigen-energy, $\psi_{nk} (\xx) = e^{i \kk \xx} u_{nk} (\xx)$ is the Bloch wavefunction, and the field operators are expanded as
\begin{equation}
\begin{aligned}
    \tilde{\psi} (\xx) = \sum_n \int d\kk \psi_{n\kk} (\xx) \hat{a}_{n\kk}, \\
    \tilde{\psi}^{\dag} (\xx) = \sum_n \int d\kk \psi_{n\kk}^* (\xx) \hat{a}_{n\kk}^{\dag},
\end{aligned}
\end{equation}
where $\hat{a}_{n\kk}$ and $\hat{a}_{n\kk}^{\dag}$ are the annihilation and creation operators, and the orthogonality of the Bloch states as well as the anticommutation relation read
\begin{equation} \label{eq:orthogonality}
    \braket{\psi_{n\kk}}{\psi_{m\kk'}} = \int d\xx \psi^*_{n\kk} (\xx) \psi_{m\kk'} (\xx) = \delta_{nm} \delta (\kk - \kk'),
\end{equation}
\begin{equation}
    \anticomm{\hat{a}_{n\kk}}{\hat{a}^{\dag}_{m\kk'}} = \delta_{nm} \delta (\kk-\kk').
\end{equation}
Next we solve the Heisenberg equation of motion of the density operator, which writes \cite{chen2022basic}
\begin{equation}
    \frac{\partial \rho_{nm}(t)}{\partial t} = -\frac{i}{\hbar} \comm{H(t)}{\rho (t)}_{nm} - \frac{\rho_{nm}(t) - \dm{0}_{nm}}{\tau},
\end{equation}
where $\dm{0}_{nm} = f_n \delta_{nm}$ is the initial density operator with the Fermi factor $f_n$ of band $n$, and the latter term is the phenomenological term 
with the relaxation time $\tau$ which describes the scattering processes of electrons \cite{chen2022basic}. We expand the density operator in powers of $\EE$ as
\begin{equation} \label{eq:dm-expansion}
    \rho (t) = \dm{0} + \dm{1} (t) + \dm{2}(t) + \cdots,
\end{equation}
and the equation of motion for the $i$-th order is then 
\begin{equation}
    \frac{\partial \dm{i}_{nm} (t)}{\partial t} = -\frac{i}{\hbar} \comm{H_0}{\dm{i}(t)}_{nm} - \frac{i}{\hbar} \comm{H_1(t)}{\dm{i-1}(t)}_{nm} - \frac{\dm{i}_{nm} (t)}{\tau_i}.
\end{equation}
The matrix element of the position operator $\xx$ is \cite{ventura2017gauge,sipe2000second,parker2019diagrammatic}
\begin{equation}
    \bra{n\kk} \xx \ket{m\kk'} = (i\partial_{\kk} + \bm{\mA}_{nm}) \delta_{nm} \delta (\kk - \kk') + \bm{\mA}_{nm} (1-\delta_{nm}) \delta (\kk - \kk'),
\end{equation}
where $\bm{\mA}_{nm} = \bra{n} i\partial_{\kk} \ket{m}$ is the Berry connection. We define the covariant derivative as follows \cite{ventura2017gauge}
\begin{equation}
    \bm{D}_{nm,\kk} = \delta_{nm} \partial_{\kk} - i \bm{\mA}_{nm},
\end{equation}
so for an operator $\mO_{nm} (k)$, the commutator with $H_1$ is 
\begin{equation}
\comm{H_1(t)}{\mO (k)}_{nm} = e \EE(t) \comm{\xx}{\mO (k)}_{nm} = i e E^b(t) \comm{D_{k^b}}{\mO (k)}_{nm},
\end{equation}
with the summation over index $b$, and
\begin{equation}
    \comm{D_{k^b}}{\mO (k)}_{nm} = \partial_b \mO_{nm} (k) - i\comm{\mA^b}{\mO (k)}_{nm},
\end{equation}
where $\partial_b = \partial/\partial k^b$. At the first order, we write $E^b(t) = E^b_{\beta} e^{-i\omega_{\beta}t} $ with summation over the frequency index $\beta$, and the equation of motion reads
\begin{equation}
\begin{aligned}
    \frac{\partial \dm{1}_{nm} (t)}{\partial t} &= -\frac{i}{\hbar} \comm{H_0}{\dm{1}(t)}_{nm} - \frac{i}{\hbar} \comm{H_1(t)}{\dm{0}}_{nm} - \frac{\dm{1}_{nm} (t)}{\tau_1} \\
    &= -\frac{i}{\hbar} \varepsilon_{nm} \dm{1}_{nm} (t) + \frac{e}{\hbar} E^b_{\beta} e^{-i\omega_{\beta}t} \comm{D^b}{\dm{0}}_{nm} - \frac{\dm{1}_{nm} (t)}{\tau_1},
\end{aligned}
\end{equation}
where $\varepsilon_{nm} = \varepsilon_n - \varepsilon_m$ with $\varepsilon_n$ being the eigen-energy of the unperturbed Hamiltonian $H_0$, and we replace $D_{k^b}$ with $D^b$.
By writing $\dm{1}_{nm} (t) = \tilde{p}^{(1)}_{nm} e^{-i\tomega_{1,nm} t}$ with $\tomega_{1,nm} = \omega_{nm} - i/\tau_1$ and $\omega_{nm} = \varepsilon_{nm}/\hbar$, we obtain
\begin{equation}
\begin{aligned}
    \frac{\partial \tilde{p}^{(1)}_{nm} }{\partial t}  = \frac{e}{\hbar} E^b_{\beta} e^{-i (\omega_{\beta} - \tomega_{1,nm} )t} \comm{D^b}{\dm{0}}_{nm}, \\
    \tilde{p}^{(1)}_{nm}  = \frac{e E^b_{\beta} e^{-i (\omega_{\beta} - \tomega_{1,nm})t} \comm{D^b}{\dm{0}}_{nm}}{i\hbar (\tomega_{1,nm}-\omega_{\beta})}.
\end{aligned}
\end{equation}
Thus, the first-order term of the density operator is
\begin{equation} \label{eq:dm1}
\begin{aligned}
    \dm{1}_{nm} (t) &= \frac{e \comm{D^b}{\dm{0}}_{nm}}{i\hbar (\omega_{nm} -\omega_{\beta} -i/\tau_1)} E^b_{\beta} e^{-i \omega_{\beta} t},
\end{aligned}
\end{equation}
where 
\begin{equation}
    \comm{D^b}{\dm{0}}_{nm} = \partial_b f_n \delta_{nm} - i f_{mn} \mA^b_{nm}.
\end{equation}
At the second order, we have 
\begin{equation}
    \frac{\partial \dm{2}_{nm} (t)}{\partial t} = -\frac{i}{\hbar} \comm{H_0}{\dm{2}(t)}_{nm} - \frac{i}{\hbar} \comm{H_1(t)}{\dm{1}}_{nm} - \frac{\dm{2}_{nm} (t)}{\tau_2}.
\end{equation}
By writing $E^c(t) = E^c_{\gamma} e^{-i\omega_{\gamma}t}$, $\dm{2}_{nm} (t) = \tilde{p}^{(2)}_{nm}  e^{-i\tomega_{2,nm} t}$ with $\tomega_{2,nm} = \omega_{nm} - i/\tau_2$, we get
\begin{equation}
    \frac{\partial \tilde{p}^{(2)}_{nm} }{\partial t}  = \frac{e}{\hbar} E^c_{\gamma} e^{-i (\omega_{\gamma} - \tomega_{2,nm})t} \comm{D^c}{\dm{1}(t)}_{nm}.
\end{equation}
We then use the expression for $\dm{1}_{nm}(t)$ in Eq. (\ref{eq:dm1}) and integrate out $t$ to obtain
\begin{equation} \label{eq:dm2}
    \dm{2}_{nm}(t) = -\frac{e^2 \Lambda^{(2)}_{nm}}{\hbar^2 (\omega_{nm} - \omega_{\Sigma} - i/\tau_2)} E^b_{\beta} E^c_{\gamma} e^{-i\omega_{\Sigma}t},
\end{equation}
where $\omega_{\Sigma} = \omega_{\beta} + \omega_{\gamma}$ and
\begin{equation} \label{eq:Gamma2}
    \Lambda^{(2)}_{nm} = \comm{D^c}{\frac{\comm{D^b}{\dm{0}}}{\tilde{\omega}_1 - \omega_{\beta}}}_{nm}.
\end{equation}
We then separate the second-order density operator $\dm{2}_{nm}(t)$ in Eq. (\ref{eq:dm2}) into the diagonal (intraband) $\dm{2}_{nn}(t)$ and off-diagonal (interband) $(1-\delta_{nm})\dm{2}_{nm}(t)$ parts \cite{sipe2000second}. The intraband contribution reads
\begin{equation}
    \dm{2}_{nn}(t) = \frac{e^2 \Lambda^{(2)}_{nn}}{\hbar^2 (\omega_{\Sigma} + i/\tau_2)} E^b_{\beta} E^c_{\gamma} e^{-i\omega_{\Sigma}t},
\end{equation}
where 
\begin{equation}
\begin{aligned}
    \Lambda^{(2)}_{nn} &= \comm{D^c}{\frac{\comm{D^b}{\dm{0}}}{\tilde{\omega}_1 - \omega_{\beta}}}_{nn} \\
    &= -\partial_c \frac{\comm{D^b}{\dm{0}}_{nn}}{ \omega_{\beta} + i/\tau_1} - i \sum_m \left(\mA^c_{nm} \frac{\comm{D^b}{\dm{0}}_{mn}}{\tilde{\omega}_{1,mn} - \omega_{\beta}} - \frac{\comm{D^b}{\dm{0}}_{nm}}{\tilde{\omega}_{1,nm} - \omega_{\beta}} \mA^c_{mn} \right) \\
    & = -\frac{\partial_b \partial_c f_n}{\omega_{\beta} + i/\tau_1} - \sum_m \left( \frac{f_{nm} \mA^b_{mn} \mA^c_{nm}}{\tilde{\omega}_{1,mn} - \omega_{\beta}} - \frac{f_{mn} \mA^b_{nm} \mA^c_{mn}}{\tilde{\omega}_{1,nm} - \omega_{\beta}} \right).
\end{aligned}
\end{equation}
The corresponding current is 
\begin{equation}
\begin{aligned}
    j^a_{\text{intra}} (t) &= -\frac{e^3}{\hbar^2 \omega_{\Sigma}} \sum_n \int_k v^a_{nn} \Lambda^{(2)}_{nn} E^b_{\beta} E^c_{\gamma} e^{-i\omega_{\Sigma}t} \\
    &= \frac{e^3}{\hbar^2 \omega_{\Sigma}} \sum_{nm} \int_k v^a_{nn} \left( \frac{f_{nm} \mA^b_{mn} \mA^c_{nm}}{\tilde{\omega}_{1,mn} - \omega_{\beta}} - \frac{f_{mn} \mA^b_{nm} \mA^c_{mn}}{\tilde{\omega}_{1,nm} - \omega_{\beta}} \right) E^b_{\beta} E^c_{\gamma} e^{-i\omega_{\Sigma}t} \\
    & = \frac{e^3}{\hbar^2 \omega_{\Sigma}}\sum_{nm} \int_k  \left( \frac{\Delta^a_{nm} f_{nm} \mA^b_{mn} \mA^c_{nm}}{\tilde{\omega}_{1,mn} - \omega_{\beta}} \right) E^b_{\beta} E^c_{\gamma} e^{-i\omega_{\Sigma}t},
\end{aligned}
\end{equation}
where $\Delta^a_{nm} = v^a_{nn} - v^a_{mm}$ and we take $\tau_2 \to \infty$ and $\partial_{\kk} f_n = 0$. We then symmetrize the indices $b\beta \leftrightarrow c\gamma$, $\kk \leftrightarrow -\kk$, and $n \leftrightarrow m$ to get
\begin{equation}
\begin{aligned}
    &j^a_{\text{intra}} (t) = \frac{e^3}{2\hbar^2 \omega_{\Sigma}}\sum_{nm} \int_k \Delta^a_{nm} f_{nm} \left( \frac{\mA^b_{mn} \mA^c_{nm}}{\tilde{\omega}_{1,mn} - \omega_{\beta}} + \frac{\mA^c_{mn} \mA^b_{nm}}{\tilde{\omega}_{1,mn} - \omega_{\gamma}} \right) E^b_{\beta} E^c_{\gamma} e^{-i\omega_{\Sigma}t} \\ 
    & =\frac{e^3}{4\hbar^2 \omega_{\Sigma}}\sum_{nm} \int_k \Delta^a_{nm} f_{nm} (\mA^b_{mn} \mA^c_{nm} -\mA^c_{mn} \mA^b_{nm} ) \left( \frac{1}{\tilde{\omega}_{1,mn} - \omega_{\beta}} - \frac{1}{\tilde{\omega}_{1,mn} - \omega_{\gamma}} \right) E^b_{\beta} E^c_{\gamma} e^{-i\omega_{\Sigma}t} \\
    & =\frac{e^3}{8\hbar^2 \omega_{\Sigma}}\sum_{nm} \int_k \Delta^a_{nm} f_{nm} (\mA^b_{mn} \mA^c_{nm} -\mA^c_{mn} \mA^b_{nm} )  \\
    &\; \; \; \times \left( \frac{1}{\tilde{\omega}_{1,mn} - \omega_{\beta}} + \frac{1}{\tilde{\omega}_{1,mn} + \omega_{\beta}} - \frac{1}{\tilde{\omega}_{1,mn} - \omega_{\gamma}} -\frac{1}{\tilde{\omega}_{1,mn} + \omega_{\gamma}}\right) E^b_{\beta} E^c_{\gamma} e^{-i\omega_{\Sigma}t}.
\end{aligned}
\end{equation}
By assuming $1/\tau_1$ is very small, i.e. $\tau_1 \to \infty$, we can apply the following identity
\begin{equation} \label{eq:identity}
    \frac{1}{\tilde{\omega}_{1,mn} - \omega_{\beta}} = \frac{1}{\omega_{mn} - \omega_{\beta} -i/\tau_1} = \frac{P}{\omega_{mn} - \omega_{\beta}} +i\pi \delta(\omega_{mn} - \omega_{\beta}),
\end{equation}
where $P$ denotes a Cauchy principle value. The imaginary part of the current is called the injection current or the circular photocurrent \cite{sipe2000second}, which reads
\begin{equation}
\begin{aligned}
    &j^a_{\text{injection}} (t)= [j^a_{\text{intra}} (t)]_{\text{Im}} = \frac{i \pi e^3}{8\hbar^2 \omega_{\Sigma}}\sum_{nm} \int_k \Delta^a_{nm} f_{nm} (\mA^b_{mn} \mA^c_{nm} -\mA^c_{mn} \mA^b_{nm} )  \\
    &\times\left[ \delta(\omega_{mn} - \omega_{\beta}) - \delta(\omega_{mn} + \omega_{\beta}) - \delta(\omega_{mn} - \omega_{\gamma}) + \delta(\omega_{mn} + \omega_{\gamma})\right] E^b_{\beta} E^c_{\gamma} e^{-i\omega_{\Sigma}t}.
\end{aligned}
\end{equation}
We then take the time derivative of the current 
\begin{equation}
\begin{aligned}
    \frac{d\, j^a_{\text{injection}} (t)}{dt} &= -i \omega_{\Sigma} \times [j^a_{\text{intra}} (t)]_{\text{Im}} \\
    &= \eta_{abc}(\omega_{\Sigma}; \omega_{\beta},\omega_{\gamma}) E^b_{\beta} E^c_{\gamma}  e^{-i\omega_{\Sigma}t},
\end{aligned}
\end{equation}
where
\begin{equation}
\begin{aligned}
    \eta_{abc}(\omega_{\Sigma}; \omega_{\beta}, \omega_{\gamma}) &= -\frac{i\pi e^3}{8\hbar^2 }\sum_{nm} \int_k \Delta^a_{nm} f_{nm} \Omega^{bc}_{mn} \\
    &\times\left[ \delta(\omega_{mn} - \omega_{\beta}) - \delta(\omega_{mn} + \omega_{\beta}) - \delta(\omega_{mn} - \omega_{\gamma}) + \delta(\omega_{mn} + \omega_{\gamma})\right].
\end{aligned}
\end{equation}
If we consider a monochromatic field $\EE(t) = \EE(\omega) e^{-i\omega t} + \EE(-\omega) e^{i\omega t}$, the injection current with $\omega_{\Sigma} = 0$ is 
\begin{equation} \label{eq:CPGE}
    \frac{d\, j^a_{\text{injection}} (t)}{dt} = 2\eta_{abc} (0; -\omega,\omega) E^b(\omega) E^c(-\omega),
\end{equation}
with
\begin{equation} \label{eq:CPGE-eta}
    \eta_{abc} (0; -\omega,\omega) = -\frac{i\pi e^3}{2\hbar^2} \sum_{nm} \int_k \Delta^a_{nm} f_{nm} \Omega^{bc}_{mn} \delta (\omega_{mn} - \omega).
\end{equation}
The same result can be also derived from the velocity gauge, as shown in Appendix \ref{sec:sm-velocity-gauge}. We can rewrite Eq. (\ref{eq:CPGE}) as \cite{de2017quantized,flicker2018chiral}
\begin{equation}
    j^a_{\text{injection}} = \tau \beta_{ab} (\omega) [\EE (\omega) \times \EE^* (\omega)]^b,
\end{equation}
with
\begin{equation}
\begin{aligned}
    \beta_{ab} (\omega) &= \frac{\pi e^3}{\hbar^2} \sum_{nm} \int \frac{d^3 k}{(2\pi)^3} (\partial_a \varepsilon_{nm}) f_{nm} R_{nm}^b \delta (\varepsilon_{mn} - \hbar \omega),
\end{aligned}
\end{equation}
where $\tau$ is the lifetime and $\bm{R}_{nm} = (\bm{\mA}_{nm} \times \bm{\mA}_{mn})$. We can see that this current can be generated by a circularly polarized light, but not a linearly polarized light, as $\EE \times \EE^*$ requires the electric fields to be along perpendicular directions, thus the name “circular photogalvanic effect”. Furthermore, the current switches direction when the polarization changes from lefthanded to righthanded. 
The CPGE trace $\beta (\omega)$ reads
\begin{equation} \label{eq:CPGE-beta}
\begin{aligned}
 \beta(\omega) = \text{Tr}[\beta_{ab}(\omega)] &= \frac{\pi e^3}{\hbar^2} \sum_{nm} \int \frac{d^3 k}{(2\pi)^3} (\partial_a \varepsilon_{nm}) f_{nm} R_{nm}^a \delta (\varepsilon_{mn} - \hbar \omega) \\
 & = \frac{e^3}{2h^2} \sum_{nm} \oint d \bm{S}_{nm} \cdot \bm{R}_{nm},
\end{aligned}
\end{equation}
where $\bm{S}_{nm}$ is a closed surface in the momentum space.
For a two-fold Weyl node with band index $n = 1,2$ and Fermi energy across at the Weyl node, we use the relation $\bm{\Omega}_{nm} = i\sum_{m\neq n} \bm{R}_{nm} $ \cite{flicker2018chiral}, and Eq. (\ref{eq:CPGE-beta}) becomes
\begin{equation}
\begin{aligned}
    \beta (\omega) & = i \frac{e^3}{2h^2} \oint d \bm{S} \cdot \bm{\Omega}_1 \\
    & = i \beta_0 C,
\end{aligned}
\end{equation}
where $C = \frac{1}{2\pi} \oint \bm{S} \cdot \bm{\Omega} $ is the Chern number and $\beta_0 = \pi e^3/ h^2$. Therefore, the CPGE trace $\beta (\omega)$ is quantized for a Weyl node as proportional to its associated topological charge, and the CPGE current can be written as
\beq
 \bm{j} = i\tau \beta_0 C \left[\EE (\omega) \times \EE^* (\omega)\right].
\eneq

For a two-band Weyl semimetal with mirror symmetry, there are two Weyl nodes with opposite topological charges $C_L = 1$ and $C_R = -1$ sitting at the same energy (Fig. \ref{fig:fig1}a). When the Fermi energy is at the Weyl node, the CPGE trace for the two Weyl nodes are $i\beta_0 C_L$ and $i\beta_0 C_R$. The CPGE currents from the left and right Weyl nodes are $ \bm{j}_L = i\tau \beta_0 C \left[\EE (\omega) \times \EE^* (\omega)\right]$ and $ \bm{j}_R = -i\tau \beta_0 C \left[\EE (\omega) \times \EE^* (\omega)\right]$, so the total CPGE current vanishes. If we consider a chiral Weyl semimetal in which the inversion and all mirror symmetries are broken (Fig. \ref{fig:fig1}b), the left and right Weyl nodes are located at different energies $\varepsilon_L$ and $\varepsilon_R$ (measured from the Fermi energy $E_f$), respectively. Then in the frequency window $2 \abs{\varepsilon_R} < \hbar \omega < 2 \abs{\varepsilon_L}$, the transition near the left Weyl node is forbidden due to Pauli blocking, and thus, the only contribution to the CPGE current comes from the right Weyl node $ \bm{j}_R = - i\tau \beta_0 C \left[\EE (\omega) \times \EE^* (\omega)\right]$ \cite{de2017quantized}. Therefore, the quantized CPGE can be realized under such conditions.
The CPGE has recently been observed in chiral Weyl semimetals RhSi \cite{rees2020helicity, ni2020linear} and CoSi \cite{ni2021giant}.

\section{Derivation of CPGE in velocity gauge} \label{sec:sm-velocity-gauge}

In this section, we derive the CPGE in the velocity gauge \cite{ventura2017gauge,parker2019diagrammatic,passos2018nonlinear}. The dipole interaction is written as
\beq
 H_0\left(\kk + \frac{e}{\hbar} \bA(t) \right) \approx H_0(\kk) + \bA (t) \cdot \JJ = H_0 (\kk) + H_1,
\eneq
where $\JJ = \frac{e}{\hbar} \frac{\partial H_0}{\partial \kk}$ is the current operator. Using the expressions in Appendix \ref{sec:sm-length-gauge}, we obtain the second-quantized Hamiltonian
\beqal
\hat{H}_0 &= \int d \xx \tpsi^{\dag} (\xx) H_0 \tpsi (\xx) \\
& = \sum_{nn'}\int d\kk d\kk' d\xx \psi^*_{n\kk} (\xx) \ha^{\dag}_{n\kk} \varepsilon_{n'\kk'} \psi_{n'\kk'} (\xx) \ha_{n'\kk'} \\
& = \sum_{n} \int d\kk \varepsilon_{n\kk} \ha^{\dag}_{n\kk} \ha_{n\kk},
\eneqal
\beqal
\hat{H}_1 & = \int d\xx \tpsi^{\dag} (\xx) \bA(t) \cdot \JJ \tpsi (\xx) \\
& = \sum_{nn'} \int d\kk d\kk' \ha^{\dag}_{n\kk} \ha_{n'\kk'} \int d\xx d\qq e^{i (-\kk + \qq + \kk')\xx} u^*_{n\kk} [\bA (\qq,t) \cdot \JJ] u_{n'\kk'}(\xx) \\
& = \sum_{nn'} \int d\kk d\kk' \ha^{\dag}_{n\kk} \ha_{n'\kk'} \int d\qq \delta (-\kk + \qq + \kk') \bra{u_{n\kk}} \bA (\qq,t) \cdot \JJ \ket{u_{n'\kk'}} \\
& = \sum_{nn'} \int d\kk d\kk' \ha^{\dag}_{n\kk} \ha_{n'\kk'} \bra{u_{n\kk}} \bA (\kk - \kk',t) \cdot \JJ \ket{u_{n'\kk'}}.
\eneqal
We assume the photon momentum is zero $\qq \approx 0$, i.e. $\kk - \kk' \approx 0$, as the momentum of light $\qq$ is much smaller than that of electrons $\kk$, so the Hamiltonian is as follows
\beq
\hat{H} = \sum_{n} \int d\kk \varepsilon_{n\kk} \ha^{\dag}_{n\kk} \ha_{n\kk} +\sum_{nn'} \int d\kk \ha^{\dag}_{n\kk} \ha_{n'\kk} \bA(t)\cdot \JJ_{nn'\kk},
\eneq
where $\JJ_{nn'\kk} = \bra{u_{n\kk}} \JJ \ket{u_{n'\kk}} $. The Heisenberg equations of motion for the creation and annihilation operators are
\beqal
\frac{\partial \ha_{n\kk}}{\partial t} &= -\frac{i}{\hbar} \comm{\hat{H}}{\ha_{n\kk}} \\
& = \frac{i}{\hbar} \left( \varepsilon_{n\kk} \ha_{n\kk} + \sum_{n'} \bA(t) \cdot \JJ_{nn'\kk} \ha_{n'\kk} \right), \\
\frac{\partial \ha^{\dag}_{n\kk}}{\partial t} &= -\frac{i}{\hbar} \comm{\hat{H}}{\ha^{\dag}_{n\kk}} \\
& = -\frac{i}{\hbar} \left( \varepsilon_{n\kk} \ha^{\dag}_{n\kk} + \sum_{n'} \bA(t) \cdot \JJ_{n'n\kk} \ha^{\dag}_{n'\kk} \right).
\eneqal
The density matrix is $\rho_{nm,\kk} = \langle \hat{a}_{m\kk}^{\dag} \hat{a}_{n\kk} \rangle$, so the optical Bloch equation for the density matrix writes
\beq
\frac{\partial \rho_{nm}(t)}{\partial t} = -\frac{i}{\hbar} \varepsilon_{nm} \rho_{nm}(t) - \frac{i}{\hbar} \sum_{n'} \left[\bA(t) \cdot \JJ_{n'n} \rho_{n'm}(t) - \bA(t) \cdot \JJ_{mn'} \rho_{nn'}(t) \right] - \frac{\rho_{nm}(t)}{\tau} .
\eneq
Next we expand the density operator in powers of $\EE$ as in Eq. (\ref{eq:dm-expansion}) and the zeroth-order term is $\dm{0}_{nm} = f_n \delta_{nm}$.
At the first order,
\beqal
\frac{\partial \dm{1}_{nm}(t)}{\partial t} &= -\frac{i}{\hbar} \varepsilon_{nm} \dm{1}_{nm}(t) - \frac{i}{\hbar} \sum_{n'} \left[\bA(t) \cdot \JJ_{n'n} \dm{0}_{n'm}(t) - \bA(t) \cdot \JJ_{mn'} \dm{0}_{nn'}(t) \right] - \frac{\dm{1}_{nm}(t)}{\tau_1} \\
&= -\frac{i}{\hbar} \varepsilon_{nm} \dm{1}_{nm}(t) - \frac{i}{\hbar} f_{mn} A^b(t) J^b_{mn} - \frac{\dm{1}_{nm}(t)}{\tau_1}.
\eneqal
Let $\dm{1}_{nm} (t) = \tp^{(1)}_{nm} e^{-i\tomega_{1,nm} t} $ with $\tomega_{1,nm} =\varepsilon_{nm}/\hbar - i/\tau_1 $, and $A^b(t) = \int \frac{d \omega_{\beta}}{2\pi} A^b (\omega_{\beta}) e^{-i\omega_{\beta} t} $, we then obtain
\beqal
 &\frac{\partial \tilde{p}^{(1)}_{nm}}{\partial t} = -\frac{i}{\hbar} \int \frac{d\omega_{\beta}}{2\pi}  f_{mn} A^b (\omega_{\beta}) J^b_{mn} e^{i(\tomega_{1,nm} -\omega_{\beta})t},  \\
 &\tp^{(1)}_{nm}= -\frac{i}{\hbar} \int \frac{d\omega_{\beta}}{2\pi} f_{mn} A^b (\omega_{\beta}) J^b_{mn} \frac{e^{i (\tomega_{1,nm} - \omega_{\beta}) t}}{i(\tomega_{1,nm} - \omega_{\beta})},
\eneqal
so the first-order density operator is
\beqal
 \dm{1}_{nm} (\omega) &= \int dt \dm{1}_{nm} (t) e^{i\omega t} \\
 &= -\frac{i}{\hbar} \int \frac{d\omega_{\beta}}{2\pi} f_{mn} A^b (\omega_{\beta}) J^b_{mn} \int dt \frac{e^{i (\omega - \omega_{\beta}) t}}{i(\tomega_{1,nm} - \omega_{\beta})} \\
 & = -\frac{i}{\hbar} \int \frac{d\omega_{\beta}}{2\pi} f_{mn} A^b (\omega_{\beta}) J^b_{mn} \frac{\delta (\omega - \omega_{\beta})}{i(\tomega_{1,nm} - \omega_{\beta})} \\
 &= \frac{f_{nm} A^b (\omega) J^b_{mn}}{\hbar (\tomega_{1,nm} - \omega)}.
\eneqal
At the second order, the diagonal part (intraband) of the density  operator reads
\beqal
\frac{\partial \dm{2}_{nn}(t)}{\partial t} = - \frac{i}{\hbar} \sum_{m} \left[\bA(t) \cdot \JJ_{mn} \dm{1}_{mn}(t) - \bA(t) \cdot \JJ_{nm} \dm{1}_{nm}(t) \right] - \frac{\dm{2}_{nn}(t)}{\tau_2}.
\eneqal
Let $\dm{2}_{nn} = e^{-t/\tau_2} \tp^{(2)}_{nn}$,
\beqal
\frac{\partial \tp^{(2)}_{nn}}{\partial t} &= -\frac{i}{\hbar} \sum_m A^c (t) [J^c_{mn} \dm{1}_{mn} (t) - J^c_{nm} \dm{1}_{nm} (t)] e^{t/\tau_2} \\
& = -\frac{i}{\hbar} \sum_m \int \frac{d\omega_{\beta} d\omega_{\gamma}}{(2\pi)^2} A^c (\omega_{\gamma}) [J^c_{mn} \dm{1}_{mn} (\omega_{\beta}) - J^c_{nm} \dm{1}_{nm} (\omega_{\beta})] e^{-i(\omega_{\beta} + \omega_{\gamma} + i/\tau_2)t}, \\
\tp^{(2)}_{nn} &= \sum_m \int \frac{d\omega_{\beta} d\omega_{\gamma}}{(2\pi)^2} A^c (\omega_{\gamma}) [J^c_{mn} \dm{1}_{mn} (\omega_{\beta}) - J^c_{nm} \dm{1}_{nm} (\omega_{\beta})] \frac{e^{-i(\omega_{\beta} + \omega_{\gamma} + i/\tau_2)t}}{\hbar (\omega_{\beta} + \omega_{\gamma} + i/\tau_2)}.
\eneqal
Thus, the second-order intraband density operator writes
\beqal
\dm{2}_{nn} (\omega) &= \int dt \tp^{(2)}_{nn} e^{i (\omega + i/\tau_2) t} \\ 
& = \sum_m \int \frac{d\omega_{\beta} d\omega_{\gamma}}{(2\pi)^2} A^c (\omega_{\gamma}) [J^c_{mn} \dm{1}_{mn} (\omega_{\beta}) - J^c_{nm} \dm{1}_{nm} (\omega_{\beta})] \frac{\delta (\omega_{\beta} + \omega_{\gamma} - \omega)}{\hbar (\omega_{\beta} + \omega_{\gamma} + i/\tau_2)} \\
& = \frac{1}{\hbar (\omega + i/\tau_2)}\sum_m \int \frac{d\omega_{\beta} d\omega_{\gamma}}{(2\pi)^2} A^c (\omega_{\gamma})[J^c_{mn} \dm{1}_{mn} (\omega_{\beta}) - J^c_{nm} \dm{1}_{nm} (\omega_{\beta})] \delta (\omega_{\beta} + \omega_{\gamma} - \omega).
\eneqal
We then use the expression for $\dm{1}_{nm}$ and $A^b_{\beta} = A^b (\omega_{\beta})$, $A^c_{\gamma} = A^c (\omega_{\gamma})$, and $\omega_{\Sigma} = \omega_{\beta} + \omega_{\gamma}$ to get
\beq
\dm{2}_{nn} (\omega_{\Sigma}) = \frac{1}{\hbar (\omega_{\Sigma} + i/\tau_2)}\sum_m A^b_{\beta} A^c_{\gamma} \left[ \frac{f_{mn}J^b_{mn} J^c_{nm} }{\hbar (\tomega_{1,mn} - \omega_{\beta})} - \frac{f_{nm}J^b_{nm} J^c_{mn} }{\hbar (\tomega_{1,nm} - \omega_{\beta})} \right],
\eneq
with the summation over indices $\beta$ and $\gamma$ implied. The corresponding current is as follows
\beqal
 j^a_{\text{intra}} (t) &= e \sum_n \int_k  v^a_{nn} \dm{2}_{nn} (\omega_{\Sigma}) e^{-i\omega_{\Sigma} t} \\
 & = \frac{e}{\hbar^2}\sum_{nm} \int_k v^a_{nn} \left[ \frac{f_{mn}J^b_{mn} J^c_{nm} }{\tomega_{1,mn} - \omega_{\beta}} - \frac{f_{nm}J^b_{nm} J^c_{mn} }{\tomega_{1,nm} - \omega_{\beta}} \right] \frac{A^b_{\beta} A^c_{\gamma} e^{-i\omega_{\Sigma} t}}{\omega_{\Sigma} + i/\tau_2} \\
 & =  \frac{e}{\hbar^2 \omega_{\Sigma}}\sum_{nm} \int_k \frac{\Delta^a_{nm} f_{mn}J^b_{mn} J^c_{nm} }{\tomega_{1,mn} - \omega_{\beta}} A^b_{\beta} A^c_{\gamma} e^{-i\omega_{\Sigma} t},
\eneqal
where $\Delta^a_{nm} = v^a_{nn} - v^a_{mm}$ is the velocity shift. We replace $A^b(\omega)$ with $E^b (\omega)/i\omega$ \cite{parker2019diagrammatic} and $J^b_{nm} = \frac{ie}{\hbar}\varepsilon_{nm} \mA^b_{nm}$ and obtain
\beq
j^a_{\text{intra}} (t) =\frac{e^3}{\hbar^2 \omega_{\Sigma}}  \sum_{nm} \int_k \frac{\Delta^a_{nm} f_{nm} \mA^b_{mn} \mA^c_{nm}}{\tomega_{1,mn} - \omega_{\beta}} \frac{\varepsilon^2_{nm} E^b_{\beta} E^c_{\gamma}}{\hbar^2 \omega_{\beta} \omega_{\gamma}} e^{-i\omega_{\Sigma} t}.
\eneq
We then take the imaginary part of the current using the identity Eq. (\ref{eq:identity}) and symmetrize all the indices to get
\beqal
 j^a _{\text{Im}} (t)&= \frac{i \pi e^3}{8\hbar^2 \omega_{\Sigma}}  \sum_{nm} \int_k \Delta^a_{nm} f_{nm} (\mA^b_{mn} \mA^c_{nm} - \mA^c_{mn} \mA^b_{nm}) \frac{\varepsilon^2_{nm} E^b_{\beta} E^c_{\gamma}}{\hbar^2 \omega_{\beta} \omega_{\gamma}} e^{-i\omega_{\Sigma} t} \\
 & \times \left[\delta(\omega_{mn} - \omega_{\beta}) - \delta(\omega_{mn} + \omega_{\beta}) - \delta(\omega_{mn} - \omega_{\gamma}) + \delta(\omega_{mn} + \omega_{\gamma}) \right].
\eneqal
The time derivative of the above current reads
\beqal
 \frac{d j^a _{\text{Im}} (t)}{dt} &= \frac{ \pi e^3}{8\hbar^2}  \sum_{nm} \int_k \Delta^a_{nm} f_{nm} (\mA^b_{mn} \mA^c_{nm} - \mA^c_{mn} \mA^b_{nm}) \frac{\varepsilon^2_{nm} E^b_{\beta} E^c_{\gamma}}{\hbar^2 \omega_{\beta} \omega_{\gamma}} e^{-i\omega_{\Sigma} t} \\
 & \times \left[\delta(\omega_{mn} - \omega_{\beta}) - \delta(\omega_{mn} + \omega_{\beta}) - \delta(\omega_{mn} - \omega_{\gamma}) + \delta(\omega_{mn} + \omega_{\gamma}) \right] \\
 & = \eta_{abc}(\omega_{\Sigma};\omega_{\beta},\omega_{\gamma} ) E^b_{\beta} E^c_{\gamma} e^{-i\omega_{\Sigma}t}.
\eneqal
For a monochromatic field with $\omega_{\beta} = -\omega_{\gamma} = \omega$ and $\omega_{\Sigma} = 0$, we have
\beqal
 \eta_{abc}(0; \omega,-\omega) &= \frac{ \pi e^3}{2\hbar^2}  \sum_{nm} \int_k \Delta^a_{nm} f_{nm} (\mA^b_{mn} \mA^c_{nm} - \mA^c_{mn} \mA^b_{nm}) \delta (\omega_{mn} - \omega) \\
 & = -\frac{i \pi e^3}{2\hbar^2}  \sum_{nm} \int_k \Delta^a_{nm} f_{nm} \Omega^{bc}_{mn} \delta (\omega_{mn} - \omega),
\eneqal
which recovers Eq. (\ref{eq:CPGE}), which is the result in the length gauge.

\section{Derivation of MR-induced nonlinear current for the four-band model} \label{sec:sm-pseudo-CPGE}

Here we drive the MR-induced nonlinear current for the four-band model Eq. (\ref{eq:four-band}). Following the same procedure in Appendix \ref{sec:sm-velocity-gauge}, we find the second-quantized Hamiltonian as follows
\beq
\hat{H} = \sum_{n} \int d\kk \varepsilon_{n\kk} \ha^{\dag}_{n\kk} \ha_{n\kk} +\sum_{nn'} \int d\kk \ha^{\dag}_{n\kk} \ha_{n'\kk} \left[\bA(t)\cdot \JJ_{nn'\kk} + \bm{\nu} (t) \cdot \bm{\Gamma}_{nn'\kk} \right],
\eneq
where $\varepsilon_{n\kk}$ is the eigenvalue of $H_0$, $\JJ$ is the current operator, and $\bm{\Gamma} = \bm{\sigma} \tau_z$. We then obtain the first-order and second-order intraband density operator
\beq
 \dm{1}_{nm} (\omega) = f_{nm} \frac{A^b (\omega) J^b_{mn} + \nu^b (\omega) \Gamma^b_{mn}}{\hbar (\tomega_{1,nm} - \omega)},
\eneq
\beqal
 \dm{2}_{nn} (\omega_{\Sigma}) = \frac{1}{\hbar \omega_{\Sigma}} \sum_m A^b_\beta \nu^c_\gamma \left[ \frac{f_{mn}J^b_{mn} \Gamma^c_{nm} }{\hbar (\tomega_{1,mn} - \omega_{\beta})} - \frac{f_{nm}J^b_{nm} \Gamma^c_{mn} }{\hbar (\tomega_{1,nm} - \omega_{\beta})} \right] \\
 + \nu^b_\beta A^c_\gamma \left[ \frac{f_{mn} \Gamma^b_{mn} J^c_{nm} }{\hbar (\tomega_{1,mn} - \omega_{\beta})} - \frac{f_{nm} \Gamma^b_{nm} J^c_{mn} }{\hbar (\tomega_{1,nm} - \omega_{\beta})} \right] .
\eneqal
The current reads
\beqal
 \tj^a (t) &= \frac{e}{\hbar^2 \omega_{\Sigma}} \sum_{nm} \int_k \frac{\Delta^a_{nm} f_{mn}}{\tomega_{1,mn} - \omega_{\beta}} \left[J^b_{mn} \Gamma^c_{nm} A^b_\beta \nu^c_\gamma + \Gamma^b_{mn} J^c_{nm} \nu^b_\beta A^c_{\gamma} \right] e^{-i\omega_{\Sigma}t} \\
 &=\frac{e^2}{\hbar^3 \omega_{\Sigma}} \sum_{nm} \int_k \frac{\Delta^a_{nm} f_{nm}}{\tomega_{1,mn} - \omega_{\beta}} \left[\frac{J^b_{mn} \Gamma^c_{nm}}{\omega_\beta \omega_\gamma} E^b_\beta \tE^c_\gamma + \frac{\Gamma^b_{mn} J^c_{nm}}{\omega_\beta \omega_\gamma} \tE^b_\beta E^c_{\gamma} \right] e^{-i\omega_{\Sigma}t},
\eneqal
where $\tE^b_\beta = i\hbar \omega_\beta \nu^b_\beta/e$ is the pseudo-electric field. Next we take the imaginary part and the time derivative of the current, and for $\omega_\beta = -\omega_\gamma = \omega$, we have
\beq
 \frac{d\tj^a (t)}{dt} = \teta_{abc} (0;\omega,-\omega) E_b (\omega) \tE_c (-\omega) + \teta_{acb}(0;-\omega,\omega) \tE_b (\omega) E_c (-\omega),
\eneq
with
\beqal
 \teta_{abc} (0;\omega,-\omega) &= \frac{\pi e^2}{\hbar^3} \sum_{nm} \int_k \Delta^a_{nm} f_{nm} \frac{J^b_{mn} \Gamma^c_{nm}}{\omega^2} \delta (\omega_{mn} - \omega) \\
 &=\frac{i\pi e^3}{\hbar^2} \sum_{nm} \int \frac{d^3 k}{(2\pi)^3} (\partial_a \varepsilon_{nm}) f_{nm} \frac{\mA^b_{nm} \Gamma^c_{mn}}{\varepsilon_{nm}} \delta (\varepsilon_{mn} - \hbar \omega),
\eneqal
where $\tilde{\EE}(\omega)$ is defined as $i\hbar \omega \bm{\nu}/e$.

\bibliography{ref.bib}

\begin{thebibliography}{50}%
\makeatletter
\providecommand \@ifxundefined [1]{%
 \@ifx{#1\undefined}
}%
\providecommand \@ifnum [1]{%
 \ifnum #1\expandafter \@firstoftwo
 \else \expandafter \@secondoftwo
 \fi
}%
\providecommand \@ifx [1]{%
 \ifx #1\expandafter \@firstoftwo
 \else \expandafter \@secondoftwo
 \fi
}%
\providecommand \natexlab [1]{#1}%
\providecommand \enquote  [1]{``#1''}%
\providecommand \bibnamefont  [1]{#1}%
\providecommand \bibfnamefont [1]{#1}%
\providecommand \citenamefont [1]{#1}%
\providecommand \href@noop [0]{\@secondoftwo}%
\providecommand \href [0]{\begingroup \@sanitize@url \@href}%
\providecommand \@href[1]{\@@startlink{#1}\@@href}%
\providecommand \@@href[1]{\endgroup#1\@@endlink}%
\providecommand \@sanitize@url [0]{\catcode `\\12\catcode `\$12\catcode `\&12\catcode `\#12\catcode `\^12\catcode `\_12\catcode `\%12\relax}%
\providecommand \@@startlink[1]{}%
\providecommand \@@endlink[0]{}%
\providecommand \url  [0]{\begingroup\@sanitize@url \@url }%
\providecommand \@url [1]{\endgroup\@href {#1}{\urlprefix }}%
\providecommand \urlprefix  [0]{URL }%
\providecommand \Eprint [0]{\href }%
\providecommand \doibase [0]{https://doi.org/}%
\providecommand \selectlanguage [0]{\@gobble}%
\providecommand \bibinfo  [0]{\@secondoftwo}%
\providecommand \bibfield  [0]{\@secondoftwo}%
\providecommand \translation [1]{[#1]}%
\providecommand \BibitemOpen [0]{}%
\providecommand \bibitemStop [0]{}%
\providecommand \bibitemNoStop [0]{.\EOS\space}%
\providecommand \EOS [0]{\spacefactor3000\relax}%
\providecommand \BibitemShut  [1]{\csname bibitem#1\endcsname}%
\let\auto@bib@innerbib\@empty
\bibitem [{\citenamefont {de~Juan}\ \emph {et~al.}(2017)\citenamefont {de~Juan}, \citenamefont {Grushin}, \citenamefont {Morimoto},\ and\ \citenamefont {Moore}}]{de2017quantized}%
  \BibitemOpen
  \bibfield  {author} {\bibinfo {author} {\bibfnamefont {F.}~\bibnamefont {de~Juan}}, \bibinfo {author} {\bibfnamefont {A.~G.}\ \bibnamefont {Grushin}}, \bibinfo {author} {\bibfnamefont {T.}~\bibnamefont {Morimoto}},\ and\ \bibinfo {author} {\bibfnamefont {J.~E.}\ \bibnamefont {Moore}},\ }\bibfield  {title} {\bibinfo {title} {Quantized circular photogalvanic effect in {W}eyl semimetals},\ }\href@noop {} {\bibfield  {journal} {\bibinfo  {journal} {Nature Communications}\ }\textbf {\bibinfo {volume} {8}},\ \bibinfo {pages} {15995} (\bibinfo {year} {2017})}\BibitemShut {NoStop}%
\bibitem [{\citenamefont {Berry}(1984)}]{berry1984quantal}%
  \BibitemOpen
  \bibfield  {author} {\bibinfo {author} {\bibfnamefont {M.~V.}\ \bibnamefont {Berry}},\ }\bibfield  {title} {\bibinfo {title} {Quantal phase factors accompanying adiabatic changes},\ }\href@noop {} {\bibfield  {journal} {\bibinfo  {journal} {Proceedings of the Royal Society of London. A. Mathematical and Physical Sciences}\ }\textbf {\bibinfo {volume} {392}},\ \bibinfo {pages} {45} (\bibinfo {year} {1984})}\BibitemShut {NoStop}%
\bibitem [{\citenamefont {Xiao}\ \emph {et~al.}(2010)\citenamefont {Xiao}, \citenamefont {Chang},\ and\ \citenamefont {Niu}}]{xiao2010berry}%
  \BibitemOpen
  \bibfield  {author} {\bibinfo {author} {\bibfnamefont {D.}~\bibnamefont {Xiao}}, \bibinfo {author} {\bibfnamefont {M.-C.}\ \bibnamefont {Chang}},\ and\ \bibinfo {author} {\bibfnamefont {Q.}~\bibnamefont {Niu}},\ }\bibfield  {title} {\bibinfo {title} {Berry phase effects on electronic properties},\ }\href@noop {} {\bibfield  {journal} {\bibinfo  {journal} {Reviews of modern physics}\ }\textbf {\bibinfo {volume} {82}},\ \bibinfo {pages} {1959} (\bibinfo {year} {2010})}\BibitemShut {NoStop}%
\bibitem [{\citenamefont {Liu}\ \emph {et~al.}(2024)\citenamefont {Liu}, \citenamefont {Qiang}, \citenamefont {Lu},\ and\ \citenamefont {Xie}}]{liu2024quantum}%
  \BibitemOpen
  \bibfield  {author} {\bibinfo {author} {\bibfnamefont {T.}~\bibnamefont {Liu}}, \bibinfo {author} {\bibfnamefont {X.-B.}\ \bibnamefont {Qiang}}, \bibinfo {author} {\bibfnamefont {H.-Z.}\ \bibnamefont {Lu}},\ and\ \bibinfo {author} {\bibfnamefont {X.}~\bibnamefont {Xie}},\ }\bibfield  {title} {\bibinfo {title} {Quantum geometry in condensed matter},\ }\href@noop {} {\bibfield  {journal} {\bibinfo  {journal} {National Science Review}\ ,\ \bibinfo {pages} {nwae334}} (\bibinfo {year} {2024})}\BibitemShut {NoStop}%
\bibitem [{\citenamefont {Resta}(2011)}]{resta2011insulating}%
  \BibitemOpen
  \bibfield  {author} {\bibinfo {author} {\bibfnamefont {R.}~\bibnamefont {Resta}},\ }\bibfield  {title} {\bibinfo {title} {The insulating state of matter: a geometrical theory},\ }\href@noop {} {\bibfield  {journal} {\bibinfo  {journal} {The European Physical Journal B}\ }\textbf {\bibinfo {volume} {79}},\ \bibinfo {pages} {121} (\bibinfo {year} {2011})}\BibitemShut {NoStop}%
\bibitem [{\citenamefont {T{\"o}rm{\"a}}(2023)}]{torma2023essay}%
  \BibitemOpen
  \bibfield  {author} {\bibinfo {author} {\bibfnamefont {P.}~\bibnamefont {T{\"o}rm{\"a}}},\ }\bibfield  {title} {\bibinfo {title} {Essay: Where can quantum geometry lead us?},\ }\href@noop {} {\bibfield  {journal} {\bibinfo  {journal} {Physical Review Letters}\ }\textbf {\bibinfo {volume} {131}},\ \bibinfo {pages} {240001} (\bibinfo {year} {2023})}\BibitemShut {NoStop}%
\bibitem [{\citenamefont {Provost}\ and\ \citenamefont {Vallee}(1980)}]{provost1980riemannian}%
  \BibitemOpen
  \bibfield  {author} {\bibinfo {author} {\bibfnamefont {J.}~\bibnamefont {Provost}}\ and\ \bibinfo {author} {\bibfnamefont {G.}~\bibnamefont {Vallee}},\ }\bibfield  {title} {\bibinfo {title} {Riemannian structure on manifolds of quantum states},\ }\href@noop {} {\bibfield  {journal} {\bibinfo  {journal} {Communications in Mathematical Physics}\ }\textbf {\bibinfo {volume} {76}},\ \bibinfo {pages} {289} (\bibinfo {year} {1980})}\BibitemShut {NoStop}%
\bibitem [{\citenamefont {Thouless}\ \emph {et~al.}(1982)\citenamefont {Thouless}, \citenamefont {Kohmoto}, \citenamefont {Nightingale},\ and\ \citenamefont {den Nijs}}]{thouless1982quantized}%
  \BibitemOpen
  \bibfield  {author} {\bibinfo {author} {\bibfnamefont {D.~J.}\ \bibnamefont {Thouless}}, \bibinfo {author} {\bibfnamefont {M.}~\bibnamefont {Kohmoto}}, \bibinfo {author} {\bibfnamefont {M.~P.}\ \bibnamefont {Nightingale}},\ and\ \bibinfo {author} {\bibfnamefont {M.}~\bibnamefont {den Nijs}},\ }\bibfield  {title} {\bibinfo {title} {Quantized {H}all conductance in a two-dimensional periodic potential},\ }\href@noop {} {\bibfield  {journal} {\bibinfo  {journal} {Physical review letters}\ }\textbf {\bibinfo {volume} {49}},\ \bibinfo {pages} {405} (\bibinfo {year} {1982})}\BibitemShut {NoStop}%
\bibitem [{\citenamefont {Sodemann}\ and\ \citenamefont {Fu}(2015)}]{sodemann2015quantum}%
  \BibitemOpen
  \bibfield  {author} {\bibinfo {author} {\bibfnamefont {I.}~\bibnamefont {Sodemann}}\ and\ \bibinfo {author} {\bibfnamefont {L.}~\bibnamefont {Fu}},\ }\bibfield  {title} {\bibinfo {title} {Quantum nonlinear {H}all effect induced by {B}erry curvature dipole in time-reversal invariant materials},\ }\href@noop {} {\bibfield  {journal} {\bibinfo  {journal} {Physical review letters}\ }\textbf {\bibinfo {volume} {115}},\ \bibinfo {pages} {216806} (\bibinfo {year} {2015})}\BibitemShut {NoStop}%
\bibitem [{\citenamefont {Du}\ \emph {et~al.}(2021)\citenamefont {Du}, \citenamefont {Lu},\ and\ \citenamefont {Xie}}]{du2021nonlinear}%
  \BibitemOpen
  \bibfield  {author} {\bibinfo {author} {\bibfnamefont {Z.}~\bibnamefont {Du}}, \bibinfo {author} {\bibfnamefont {H.-Z.}\ \bibnamefont {Lu}},\ and\ \bibinfo {author} {\bibfnamefont {X.}~\bibnamefont {Xie}},\ }\bibfield  {title} {\bibinfo {title} {Nonlinear {H}all effects},\ }\href@noop {} {\bibfield  {journal} {\bibinfo  {journal} {Nature Reviews Physics}\ }\textbf {\bibinfo {volume} {3}},\ \bibinfo {pages} {744} (\bibinfo {year} {2021})}\BibitemShut {NoStop}%
\bibitem [{\citenamefont {Zhang}\ \emph {et~al.}(2018{\natexlab{a}})\citenamefont {Zhang}, \citenamefont {Sun},\ and\ \citenamefont {Yan}}]{zhang2018berry}%
  \BibitemOpen
  \bibfield  {author} {\bibinfo {author} {\bibfnamefont {Y.}~\bibnamefont {Zhang}}, \bibinfo {author} {\bibfnamefont {Y.}~\bibnamefont {Sun}},\ and\ \bibinfo {author} {\bibfnamefont {B.}~\bibnamefont {Yan}},\ }\bibfield  {title} {\bibinfo {title} {Berry curvature dipole in {W}eyl semimetal materials: an ab initio study},\ }\href@noop {} {\bibfield  {journal} {\bibinfo  {journal} {Physical Review B}\ }\textbf {\bibinfo {volume} {97}},\ \bibinfo {pages} {041101} (\bibinfo {year} {2018}{\natexlab{a}})}\BibitemShut {NoStop}%
\bibitem [{\citenamefont {Wang}\ \emph {et~al.}(2021)\citenamefont {Wang}, \citenamefont {Gao},\ and\ \citenamefont {Xiao}}]{wang2021intrinsic}%
  \BibitemOpen
  \bibfield  {author} {\bibinfo {author} {\bibfnamefont {C.}~\bibnamefont {Wang}}, \bibinfo {author} {\bibfnamefont {Y.}~\bibnamefont {Gao}},\ and\ \bibinfo {author} {\bibfnamefont {D.}~\bibnamefont {Xiao}},\ }\bibfield  {title} {\bibinfo {title} {Intrinsic nonlinear hall effect in antiferromagnetic tetragonal cumnas},\ }\href@noop {} {\bibfield  {journal} {\bibinfo  {journal} {Physical Review Letters}\ }\textbf {\bibinfo {volume} {127}},\ \bibinfo {pages} {277201} (\bibinfo {year} {2021})}\BibitemShut {NoStop}%
\bibitem [{\citenamefont {Das}\ \emph {et~al.}(2023)\citenamefont {Das}, \citenamefont {Lahiri}, \citenamefont {Atencia}, \citenamefont {Culcer},\ and\ \citenamefont {Agarwal}}]{das2023intrinsic}%
  \BibitemOpen
  \bibfield  {author} {\bibinfo {author} {\bibfnamefont {K.}~\bibnamefont {Das}}, \bibinfo {author} {\bibfnamefont {S.}~\bibnamefont {Lahiri}}, \bibinfo {author} {\bibfnamefont {R.~B.}\ \bibnamefont {Atencia}}, \bibinfo {author} {\bibfnamefont {D.}~\bibnamefont {Culcer}},\ and\ \bibinfo {author} {\bibfnamefont {A.}~\bibnamefont {Agarwal}},\ }\bibfield  {title} {\bibinfo {title} {Intrinsic nonlinear conductivities induced by the quantum metric},\ }\href@noop {} {\bibfield  {journal} {\bibinfo  {journal} {Physical Review B}\ }\textbf {\bibinfo {volume} {108}},\ \bibinfo {pages} {L201405} (\bibinfo {year} {2023})}\BibitemShut {NoStop}%
\bibitem [{\citenamefont {Kaplan}\ \emph {et~al.}(2024)\citenamefont {Kaplan}, \citenamefont {Holder},\ and\ \citenamefont {Yan}}]{kaplan2024unification}%
  \BibitemOpen
  \bibfield  {author} {\bibinfo {author} {\bibfnamefont {D.}~\bibnamefont {Kaplan}}, \bibinfo {author} {\bibfnamefont {T.}~\bibnamefont {Holder}},\ and\ \bibinfo {author} {\bibfnamefont {B.}~\bibnamefont {Yan}},\ }\bibfield  {title} {\bibinfo {title} {Unification of nonlinear anomalous {H}all effect and nonreciprocal magnetoresistance in metals by the quantum geometry},\ }\href@noop {} {\bibfield  {journal} {\bibinfo  {journal} {Physical review letters}\ }\textbf {\bibinfo {volume} {132}},\ \bibinfo {pages} {026301} (\bibinfo {year} {2024})}\BibitemShut {NoStop}%
\bibitem [{\citenamefont {Ma}\ \emph {et~al.}(2019)\citenamefont {Ma}, \citenamefont {Xu}, \citenamefont {Shen}, \citenamefont {MacNeill}, \citenamefont {Fatemi}, \citenamefont {Chang}, \citenamefont {Mier~Valdivia}, \citenamefont {Wu}, \citenamefont {Du}, \citenamefont {Hsu} \emph {et~al.}}]{ma2019observation}%
  \BibitemOpen
  \bibfield  {author} {\bibinfo {author} {\bibfnamefont {Q.}~\bibnamefont {Ma}}, \bibinfo {author} {\bibfnamefont {S.-Y.}\ \bibnamefont {Xu}}, \bibinfo {author} {\bibfnamefont {H.}~\bibnamefont {Shen}}, \bibinfo {author} {\bibfnamefont {D.}~\bibnamefont {MacNeill}}, \bibinfo {author} {\bibfnamefont {V.}~\bibnamefont {Fatemi}}, \bibinfo {author} {\bibfnamefont {T.-R.}\ \bibnamefont {Chang}}, \bibinfo {author} {\bibfnamefont {A.~M.}\ \bibnamefont {Mier~Valdivia}}, \bibinfo {author} {\bibfnamefont {S.}~\bibnamefont {Wu}}, \bibinfo {author} {\bibfnamefont {Z.}~\bibnamefont {Du}}, \bibinfo {author} {\bibfnamefont {C.-H.}\ \bibnamefont {Hsu}}, \emph {et~al.},\ }\bibfield  {title} {\bibinfo {title} {Observation of the nonlinear {H}all effect under time-reversal-symmetric conditions},\ }\href@noop {} {\bibfield  {journal} {\bibinfo  {journal} {Nature}\ }\textbf {\bibinfo {volume} {565}},\ \bibinfo {pages} {337} (\bibinfo {year} {2019})}\BibitemShut {NoStop}%
\bibitem [{\citenamefont {Kang}\ \emph {et~al.}(2019)\citenamefont {Kang}, \citenamefont {Li}, \citenamefont {Sohn}, \citenamefont {Shan},\ and\ \citenamefont {Mak}}]{kang2019nonlinear}%
  \BibitemOpen
  \bibfield  {author} {\bibinfo {author} {\bibfnamefont {K.}~\bibnamefont {Kang}}, \bibinfo {author} {\bibfnamefont {T.}~\bibnamefont {Li}}, \bibinfo {author} {\bibfnamefont {E.}~\bibnamefont {Sohn}}, \bibinfo {author} {\bibfnamefont {J.}~\bibnamefont {Shan}},\ and\ \bibinfo {author} {\bibfnamefont {K.~F.}\ \bibnamefont {Mak}},\ }\bibfield  {title} {\bibinfo {title} {Nonlinear anomalous {H}all effect in few-layer {WT}e$_2$},\ }\href@noop {} {\bibfield  {journal} {\bibinfo  {journal} {Nature materials}\ }\textbf {\bibinfo {volume} {18}},\ \bibinfo {pages} {324} (\bibinfo {year} {2019})}\BibitemShut {NoStop}%
\bibitem [{\citenamefont {Shvetsov}\ \emph {et~al.}(2019)\citenamefont {Shvetsov}, \citenamefont {Esin}, \citenamefont {Timonina}, \citenamefont {Kolesnikov},\ and\ \citenamefont {Deviatov}}]{shvetsov2019nonlinear}%
  \BibitemOpen
  \bibfield  {author} {\bibinfo {author} {\bibfnamefont {O.~O.}\ \bibnamefont {Shvetsov}}, \bibinfo {author} {\bibfnamefont {V.~D.}\ \bibnamefont {Esin}}, \bibinfo {author} {\bibfnamefont {A.~V.}\ \bibnamefont {Timonina}}, \bibinfo {author} {\bibfnamefont {N.~N.}\ \bibnamefont {Kolesnikov}},\ and\ \bibinfo {author} {\bibfnamefont {E.}~\bibnamefont {Deviatov}},\ }\bibfield  {title} {\bibinfo {title} {Nonlinear {H}all effect in three-dimensional {W}eyl and {D}irac semimetals},\ }\href@noop {} {\bibfield  {journal} {\bibinfo  {journal} {JETP Letters}\ }\textbf {\bibinfo {volume} {109}},\ \bibinfo {pages} {715} (\bibinfo {year} {2019})}\BibitemShut {NoStop}%
\bibitem [{\citenamefont {Dzsaber}\ \emph {et~al.}(2021)\citenamefont {Dzsaber}, \citenamefont {Yan}, \citenamefont {Taupin}, \citenamefont {Eguchi}, \citenamefont {Prokofiev}, \citenamefont {Shiroka}, \citenamefont {Blaha}, \citenamefont {Rubel}, \citenamefont {Grefe}, \citenamefont {Lai} \emph {et~al.}}]{dzsaber2021giant}%
  \BibitemOpen
  \bibfield  {author} {\bibinfo {author} {\bibfnamefont {S.}~\bibnamefont {Dzsaber}}, \bibinfo {author} {\bibfnamefont {X.}~\bibnamefont {Yan}}, \bibinfo {author} {\bibfnamefont {M.}~\bibnamefont {Taupin}}, \bibinfo {author} {\bibfnamefont {G.}~\bibnamefont {Eguchi}}, \bibinfo {author} {\bibfnamefont {A.}~\bibnamefont {Prokofiev}}, \bibinfo {author} {\bibfnamefont {T.}~\bibnamefont {Shiroka}}, \bibinfo {author} {\bibfnamefont {P.}~\bibnamefont {Blaha}}, \bibinfo {author} {\bibfnamefont {O.}~\bibnamefont {Rubel}}, \bibinfo {author} {\bibfnamefont {S.~E.}\ \bibnamefont {Grefe}}, \bibinfo {author} {\bibfnamefont {H.-H.}\ \bibnamefont {Lai}}, \emph {et~al.},\ }\bibfield  {title} {\bibinfo {title} {Giant spontaneous {H}all effect in a nonmagnetic {W}eyl--{K}ondo semimetal},\ }\href@noop {} {\bibfield  {journal} {\bibinfo  {journal} {Proceedings of the National Academy of Sciences}\ }\textbf {\bibinfo {volume} {118}},\ \bibinfo {pages} {e2013386118} (\bibinfo {year} {2021})}\BibitemShut {NoStop}%
\bibitem [{\citenamefont {Qin}\ \emph {et~al.}(2021)\citenamefont {Qin}, \citenamefont {Zhu}, \citenamefont {Ye}, \citenamefont {Xu}, \citenamefont {Song}, \citenamefont {Liang}, \citenamefont {Liu},\ and\ \citenamefont {Liao}}]{qin2021strain}%
  \BibitemOpen
  \bibfield  {author} {\bibinfo {author} {\bibfnamefont {M.-S.}\ \bibnamefont {Qin}}, \bibinfo {author} {\bibfnamefont {P.-F.}\ \bibnamefont {Zhu}}, \bibinfo {author} {\bibfnamefont {X.-G.}\ \bibnamefont {Ye}}, \bibinfo {author} {\bibfnamefont {W.-Z.}\ \bibnamefont {Xu}}, \bibinfo {author} {\bibfnamefont {Z.-H.}\ \bibnamefont {Song}}, \bibinfo {author} {\bibfnamefont {J.}~\bibnamefont {Liang}}, \bibinfo {author} {\bibfnamefont {K.}~\bibnamefont {Liu}},\ and\ \bibinfo {author} {\bibfnamefont {Z.-M.}\ \bibnamefont {Liao}},\ }\bibfield  {title} {\bibinfo {title} {Strain tunable {B}erry curvature dipole, orbital magnetization and nonlinear {H}all effect in {WS}e$_2$ monolayer},\ }\href@noop {} {\bibfield  {journal} {\bibinfo  {journal} {Chinese Physics Letters}\ }\textbf {\bibinfo {volume} {38}},\ \bibinfo {pages} {017301} (\bibinfo {year} {2021})}\BibitemShut {NoStop}%
\bibitem [{\citenamefont {Tiwari}\ \emph {et~al.}(2021)\citenamefont {Tiwari}, \citenamefont {Chen}, \citenamefont {Zhong}, \citenamefont {Drueke}, \citenamefont {Koo}, \citenamefont {Kaczmarek}, \citenamefont {Xiao}, \citenamefont {Gao}, \citenamefont {Luo}, \citenamefont {Niu} \emph {et~al.}}]{tiwari2021giant}%
  \BibitemOpen
  \bibfield  {author} {\bibinfo {author} {\bibfnamefont {A.}~\bibnamefont {Tiwari}}, \bibinfo {author} {\bibfnamefont {F.}~\bibnamefont {Chen}}, \bibinfo {author} {\bibfnamefont {S.}~\bibnamefont {Zhong}}, \bibinfo {author} {\bibfnamefont {E.}~\bibnamefont {Drueke}}, \bibinfo {author} {\bibfnamefont {J.}~\bibnamefont {Koo}}, \bibinfo {author} {\bibfnamefont {A.}~\bibnamefont {Kaczmarek}}, \bibinfo {author} {\bibfnamefont {C.}~\bibnamefont {Xiao}}, \bibinfo {author} {\bibfnamefont {J.}~\bibnamefont {Gao}}, \bibinfo {author} {\bibfnamefont {X.}~\bibnamefont {Luo}}, \bibinfo {author} {\bibfnamefont {Q.}~\bibnamefont {Niu}}, \emph {et~al.},\ }\bibfield  {title} {\bibinfo {title} {Giant c-axis nonlinear anomalous hall effect in {T}d-{M}o{T}e$_2$ and {WT}e$_2$},\ }\href@noop {} {\bibfield  {journal} {\bibinfo  {journal} {Nature communications}\ }\textbf {\bibinfo {volume} {12}},\ \bibinfo {pages} {2049} (\bibinfo {year} {2021})}\BibitemShut {NoStop}%
\bibitem [{\citenamefont {Ma}\ \emph {et~al.}(2022)\citenamefont {Ma}, \citenamefont {Chen}, \citenamefont {Yananose}, \citenamefont {Zhou}, \citenamefont {Wang}, \citenamefont {Li}, \citenamefont {Zhu}, \citenamefont {Wu}, \citenamefont {Xu}, \citenamefont {Yu} \emph {et~al.}}]{ma2022growth}%
  \BibitemOpen
  \bibfield  {author} {\bibinfo {author} {\bibfnamefont {T.}~\bibnamefont {Ma}}, \bibinfo {author} {\bibfnamefont {H.}~\bibnamefont {Chen}}, \bibinfo {author} {\bibfnamefont {K.}~\bibnamefont {Yananose}}, \bibinfo {author} {\bibfnamefont {X.}~\bibnamefont {Zhou}}, \bibinfo {author} {\bibfnamefont {L.}~\bibnamefont {Wang}}, \bibinfo {author} {\bibfnamefont {R.}~\bibnamefont {Li}}, \bibinfo {author} {\bibfnamefont {Z.}~\bibnamefont {Zhu}}, \bibinfo {author} {\bibfnamefont {Z.}~\bibnamefont {Wu}}, \bibinfo {author} {\bibfnamefont {Q.-H.}\ \bibnamefont {Xu}}, \bibinfo {author} {\bibfnamefont {J.}~\bibnamefont {Yu}}, \emph {et~al.},\ }\bibfield  {title} {\bibinfo {title} {Growth of bilayer {M}o{T}e$_2$ single crystals with strong non-linear {H}all effect},\ }\href@noop {} {\bibfield  {journal} {\bibinfo  {journal} {Nature communications}\ }\textbf {\bibinfo {volume} {13}},\ \bibinfo {pages} {5465} (\bibinfo {year} {2022})}\BibitemShut {NoStop}%
\bibitem [{\citenamefont {Huang}\ \emph {et~al.}(2023{\natexlab{a}})\citenamefont {Huang}, \citenamefont {Wu}, \citenamefont {Zhang}, \citenamefont {Feng}, \citenamefont {Zhou}, \citenamefont {Wang}, \citenamefont {Chen}, \citenamefont {Cheng}, \citenamefont {Sun}, \citenamefont {Meng} \emph {et~al.}}]{huang2023intrinsic}%
  \BibitemOpen
  \bibfield  {author} {\bibinfo {author} {\bibfnamefont {M.}~\bibnamefont {Huang}}, \bibinfo {author} {\bibfnamefont {Z.}~\bibnamefont {Wu}}, \bibinfo {author} {\bibfnamefont {X.}~\bibnamefont {Zhang}}, \bibinfo {author} {\bibfnamefont {X.}~\bibnamefont {Feng}}, \bibinfo {author} {\bibfnamefont {Z.}~\bibnamefont {Zhou}}, \bibinfo {author} {\bibfnamefont {S.}~\bibnamefont {Wang}}, \bibinfo {author} {\bibfnamefont {Y.}~\bibnamefont {Chen}}, \bibinfo {author} {\bibfnamefont {C.}~\bibnamefont {Cheng}}, \bibinfo {author} {\bibfnamefont {K.}~\bibnamefont {Sun}}, \bibinfo {author} {\bibfnamefont {Z.~Y.}\ \bibnamefont {Meng}}, \emph {et~al.},\ }\bibfield  {title} {\bibinfo {title} {Intrinsic nonlinear {H}all effect and gate-switchable {B}erry curvature sliding in twisted bilayer graphene},\ }\href@noop {} {\bibfield  {journal} {\bibinfo  {journal} {Physical Review Letters}\ }\textbf {\bibinfo {volume} {131}},\ \bibinfo {pages} {066301} (\bibinfo {year} {2023}{\natexlab{a}})}\BibitemShut {NoStop}%
\bibitem [{\citenamefont {Huang}\ \emph {et~al.}(2023{\natexlab{b}})\citenamefont {Huang}, \citenamefont {Wu}, \citenamefont {Hu}, \citenamefont {Cai}, \citenamefont {Li}, \citenamefont {An}, \citenamefont {Feng}, \citenamefont {Ye}, \citenamefont {Lin}, \citenamefont {Law} \emph {et~al.}}]{huang2023giant}%
  \BibitemOpen
  \bibfield  {author} {\bibinfo {author} {\bibfnamefont {M.}~\bibnamefont {Huang}}, \bibinfo {author} {\bibfnamefont {Z.}~\bibnamefont {Wu}}, \bibinfo {author} {\bibfnamefont {J.}~\bibnamefont {Hu}}, \bibinfo {author} {\bibfnamefont {X.}~\bibnamefont {Cai}}, \bibinfo {author} {\bibfnamefont {E.}~\bibnamefont {Li}}, \bibinfo {author} {\bibfnamefont {L.}~\bibnamefont {An}}, \bibinfo {author} {\bibfnamefont {X.}~\bibnamefont {Feng}}, \bibinfo {author} {\bibfnamefont {Z.}~\bibnamefont {Ye}}, \bibinfo {author} {\bibfnamefont {N.}~\bibnamefont {Lin}}, \bibinfo {author} {\bibfnamefont {K.~T.}\ \bibnamefont {Law}}, \emph {et~al.},\ }\bibfield  {title} {\bibinfo {title} {Giant nonlinear {H}all effect in twisted bilayer {WS}e$_2$},\ }\href@noop {} {\bibfield  {journal} {\bibinfo  {journal} {National Science Review}\ }\textbf {\bibinfo {volume} {10}},\ \bibinfo {pages} {nwac232} (\bibinfo {year} {2023}{\natexlab{b}})}\BibitemShut {NoStop}%
\bibitem [{\citenamefont {Zhao}\ \emph {et~al.}(2023)\citenamefont {Zhao}, \citenamefont {Cao}, \citenamefont {Zhang}, \citenamefont {Li}, \citenamefont {Li}, \citenamefont {Ma},\ and\ \citenamefont {Yang}}]{zhao2023berry}%
  \BibitemOpen
  \bibfield  {author} {\bibinfo {author} {\bibfnamefont {Y.}~\bibnamefont {Zhao}}, \bibinfo {author} {\bibfnamefont {J.}~\bibnamefont {Cao}}, \bibinfo {author} {\bibfnamefont {Z.}~\bibnamefont {Zhang}}, \bibinfo {author} {\bibfnamefont {S.}~\bibnamefont {Li}}, \bibinfo {author} {\bibfnamefont {Y.}~\bibnamefont {Li}}, \bibinfo {author} {\bibfnamefont {F.}~\bibnamefont {Ma}},\ and\ \bibinfo {author} {\bibfnamefont {S.~A.}\ \bibnamefont {Yang}},\ }\bibfield  {title} {\bibinfo {title} {Berry curvature dipole and nonlinear {H}all effect in two-dimensional {N}b$_{2n+1}${S}i$_n${T}e$_{4n+2}$},\ }\href@noop {} {\bibfield  {journal} {\bibinfo  {journal} {Physical Review B}\ }\textbf {\bibinfo {volume} {107}},\ \bibinfo {pages} {205124} (\bibinfo {year} {2023})}\BibitemShut {NoStop}%
\bibitem [{\citenamefont {Gao}\ \emph {et~al.}(2023)\citenamefont {Gao}, \citenamefont {Liu}, \citenamefont {Qiu}, \citenamefont {Ghosh}, \citenamefont {V.~Trevisan}, \citenamefont {Onishi}, \citenamefont {Hu}, \citenamefont {Qian}, \citenamefont {Tien}, \citenamefont {Chen} \emph {et~al.}}]{gao2023quantum}%
  \BibitemOpen
  \bibfield  {author} {\bibinfo {author} {\bibfnamefont {A.}~\bibnamefont {Gao}}, \bibinfo {author} {\bibfnamefont {Y.-F.}\ \bibnamefont {Liu}}, \bibinfo {author} {\bibfnamefont {J.-X.}\ \bibnamefont {Qiu}}, \bibinfo {author} {\bibfnamefont {B.}~\bibnamefont {Ghosh}}, \bibinfo {author} {\bibfnamefont {T.}~\bibnamefont {V.~Trevisan}}, \bibinfo {author} {\bibfnamefont {Y.}~\bibnamefont {Onishi}}, \bibinfo {author} {\bibfnamefont {C.}~\bibnamefont {Hu}}, \bibinfo {author} {\bibfnamefont {T.}~\bibnamefont {Qian}}, \bibinfo {author} {\bibfnamefont {H.-J.}\ \bibnamefont {Tien}}, \bibinfo {author} {\bibfnamefont {S.-W.}\ \bibnamefont {Chen}}, \emph {et~al.},\ }\bibfield  {title} {\bibinfo {title} {Quantum metric nonlinear {H}all effect in a topological antiferromagnetic heterostructure},\ }\href@noop {} {\bibfield  {journal} {\bibinfo  {journal} {Science}\ }\textbf {\bibinfo {volume} {381}},\ \bibinfo {pages} {181} (\bibinfo {year} {2023})}\BibitemShut {NoStop}%
\bibitem [{\citenamefont {Wang}\ \emph {et~al.}(2023)\citenamefont {Wang}, \citenamefont {Kaplan}, \citenamefont {Zhang}, \citenamefont {Holder}, \citenamefont {Cao}, \citenamefont {Wang}, \citenamefont {Zhou}, \citenamefont {Zhou}, \citenamefont {Jiang}, \citenamefont {Zhang} \emph {et~al.}}]{wang2023quantum}%
  \BibitemOpen
  \bibfield  {author} {\bibinfo {author} {\bibfnamefont {N.}~\bibnamefont {Wang}}, \bibinfo {author} {\bibfnamefont {D.}~\bibnamefont {Kaplan}}, \bibinfo {author} {\bibfnamefont {Z.}~\bibnamefont {Zhang}}, \bibinfo {author} {\bibfnamefont {T.}~\bibnamefont {Holder}}, \bibinfo {author} {\bibfnamefont {N.}~\bibnamefont {Cao}}, \bibinfo {author} {\bibfnamefont {A.}~\bibnamefont {Wang}}, \bibinfo {author} {\bibfnamefont {X.}~\bibnamefont {Zhou}}, \bibinfo {author} {\bibfnamefont {F.}~\bibnamefont {Zhou}}, \bibinfo {author} {\bibfnamefont {Z.}~\bibnamefont {Jiang}}, \bibinfo {author} {\bibfnamefont {C.}~\bibnamefont {Zhang}}, \emph {et~al.},\ }\bibfield  {title} {\bibinfo {title} {Quantum-metric-induced nonlinear transport in a topological antiferromagnet},\ }\href@noop {} {\bibfield  {journal} {\bibinfo  {journal} {Nature}\ }\textbf {\bibinfo {volume} {621}},\ \bibinfo {pages} {487} (\bibinfo {year} {2023})}\BibitemShut {NoStop}%
\bibitem [{\citenamefont {Le}\ \emph {et~al.}(2020)\citenamefont {Le}, \citenamefont {Zhang}, \citenamefont {Felser},\ and\ \citenamefont {Sun}}]{le2020ab}%
  \BibitemOpen
  \bibfield  {author} {\bibinfo {author} {\bibfnamefont {C.}~\bibnamefont {Le}}, \bibinfo {author} {\bibfnamefont {Y.}~\bibnamefont {Zhang}}, \bibinfo {author} {\bibfnamefont {C.}~\bibnamefont {Felser}},\ and\ \bibinfo {author} {\bibfnamefont {Y.}~\bibnamefont {Sun}},\ }\bibfield  {title} {\bibinfo {title} {Ab initio study of quantized circular photogalvanic effect in chiral multifold semimetals},\ }\href@noop {} {\bibfield  {journal} {\bibinfo  {journal} {Physical Review B}\ }\textbf {\bibinfo {volume} {102}},\ \bibinfo {pages} {121111} (\bibinfo {year} {2020})}\BibitemShut {NoStop}%
\bibitem [{\citenamefont {Le}\ and\ \citenamefont {Sun}(2021)}]{le2021topology}%
  \BibitemOpen
  \bibfield  {author} {\bibinfo {author} {\bibfnamefont {C.}~\bibnamefont {Le}}\ and\ \bibinfo {author} {\bibfnamefont {Y.}~\bibnamefont {Sun}},\ }\bibfield  {title} {\bibinfo {title} {Topology and symmetry of circular photogalvanic effect in the chiral multifold semimetals: a review},\ }\href@noop {} {\bibfield  {journal} {\bibinfo  {journal} {Journal of Physics: Condensed Matter}\ }\textbf {\bibinfo {volume} {33}},\ \bibinfo {pages} {503003} (\bibinfo {year} {2021})}\BibitemShut {NoStop}%
\bibitem [{\citenamefont {Rees}\ \emph {et~al.}(2020)\citenamefont {Rees}, \citenamefont {Manna}, \citenamefont {Lu}, \citenamefont {Morimoto}, \citenamefont {Borrmann}, \citenamefont {Felser}, \citenamefont {Moore}, \citenamefont {Torchinsky},\ and\ \citenamefont {Orenstein}}]{rees2020helicity}%
  \BibitemOpen
  \bibfield  {author} {\bibinfo {author} {\bibfnamefont {D.}~\bibnamefont {Rees}}, \bibinfo {author} {\bibfnamefont {K.}~\bibnamefont {Manna}}, \bibinfo {author} {\bibfnamefont {B.}~\bibnamefont {Lu}}, \bibinfo {author} {\bibfnamefont {T.}~\bibnamefont {Morimoto}}, \bibinfo {author} {\bibfnamefont {H.}~\bibnamefont {Borrmann}}, \bibinfo {author} {\bibfnamefont {C.}~\bibnamefont {Felser}}, \bibinfo {author} {\bibfnamefont {J.}~\bibnamefont {Moore}}, \bibinfo {author} {\bibfnamefont {D.~H.}\ \bibnamefont {Torchinsky}},\ and\ \bibinfo {author} {\bibfnamefont {J.}~\bibnamefont {Orenstein}},\ }\bibfield  {title} {\bibinfo {title} {Helicity-dependent photocurrents in the chiral weyl semimetal {R}h{S}i},\ }\href@noop {} {\bibfield  {journal} {\bibinfo  {journal} {Science advances}\ }\textbf {\bibinfo {volume} {6}},\ \bibinfo {pages} {eaba0509} (\bibinfo {year} {2020})}\BibitemShut {NoStop}%
\bibitem [{\citenamefont {Ni}\ \emph {et~al.}(2021)\citenamefont {Ni}, \citenamefont {Wang}, \citenamefont {Zhang}, \citenamefont {Pozo}, \citenamefont {Xu}, \citenamefont {Han}, \citenamefont {Manna}, \citenamefont {Paglione}, \citenamefont {Felser}, \citenamefont {Grushin} \emph {et~al.}}]{ni2021giant}%
  \BibitemOpen
  \bibfield  {author} {\bibinfo {author} {\bibfnamefont {Z.}~\bibnamefont {Ni}}, \bibinfo {author} {\bibfnamefont {K.}~\bibnamefont {Wang}}, \bibinfo {author} {\bibfnamefont {Y.}~\bibnamefont {Zhang}}, \bibinfo {author} {\bibfnamefont {O.}~\bibnamefont {Pozo}}, \bibinfo {author} {\bibfnamefont {B.}~\bibnamefont {Xu}}, \bibinfo {author} {\bibfnamefont {X.}~\bibnamefont {Han}}, \bibinfo {author} {\bibfnamefont {K.}~\bibnamefont {Manna}}, \bibinfo {author} {\bibfnamefont {J.}~\bibnamefont {Paglione}}, \bibinfo {author} {\bibfnamefont {C.}~\bibnamefont {Felser}}, \bibinfo {author} {\bibfnamefont {A.~G.}\ \bibnamefont {Grushin}}, \emph {et~al.},\ }\bibfield  {title} {\bibinfo {title} {Giant topological longitudinal circular photo-galvanic effect in the chiral multifold semimetal {C}o{S}i},\ }\href@noop {} {\bibfield  {journal} {\bibinfo  {journal} {Nature communications}\ }\textbf {\bibinfo {volume} {12}},\ \bibinfo {pages} {154} (\bibinfo {year} {2021})}\BibitemShut {NoStop}%
\bibitem [{\citenamefont {Ni}\ \emph {et~al.}(2020)\citenamefont {Ni}, \citenamefont {Xu}, \citenamefont {S{\'a}nchez-Mart{\'\i}nez}, \citenamefont {Zhang}, \citenamefont {Manna}, \citenamefont {Bernhard}, \citenamefont {Venderbos}, \citenamefont {De~Juan}, \citenamefont {Felser}, \citenamefont {Grushin} \emph {et~al.}}]{ni2020linear}%
  \BibitemOpen
  \bibfield  {author} {\bibinfo {author} {\bibfnamefont {Z.}~\bibnamefont {Ni}}, \bibinfo {author} {\bibfnamefont {B.}~\bibnamefont {Xu}}, \bibinfo {author} {\bibfnamefont {M.-{\'A}.}\ \bibnamefont {S{\'a}nchez-Mart{\'\i}nez}}, \bibinfo {author} {\bibfnamefont {Y.}~\bibnamefont {Zhang}}, \bibinfo {author} {\bibfnamefont {K.}~\bibnamefont {Manna}}, \bibinfo {author} {\bibfnamefont {C.}~\bibnamefont {Bernhard}}, \bibinfo {author} {\bibfnamefont {J.}~\bibnamefont {Venderbos}}, \bibinfo {author} {\bibfnamefont {F.}~\bibnamefont {De~Juan}}, \bibinfo {author} {\bibfnamefont {C.}~\bibnamefont {Felser}}, \bibinfo {author} {\bibfnamefont {A.~G.}\ \bibnamefont {Grushin}}, \emph {et~al.},\ }\bibfield  {title} {\bibinfo {title} {Linear and nonlinear optical responses in the chiral multifold semimetal {R}h{S}i},\ }\href@noop {} {\bibfield  {journal} {\bibinfo  {journal} {npj Quantum Materials}\ }\textbf {\bibinfo {volume} {5}},\ \bibinfo {pages} {96} (\bibinfo {year} {2020})}\BibitemShut {NoStop}%
\bibitem [{\citenamefont {Ji}\ \emph {et~al.}(2019)\citenamefont {Ji}, \citenamefont {Liu}, \citenamefont {Addison}, \citenamefont {Liu}, \citenamefont {Yu}, \citenamefont {Gao}, \citenamefont {Liu}, \citenamefont {Rappe}, \citenamefont {Kane}, \citenamefont {Mele} \emph {et~al.}}]{ji2019spatially}%
  \BibitemOpen
  \bibfield  {author} {\bibinfo {author} {\bibfnamefont {Z.}~\bibnamefont {Ji}}, \bibinfo {author} {\bibfnamefont {G.}~\bibnamefont {Liu}}, \bibinfo {author} {\bibfnamefont {Z.}~\bibnamefont {Addison}}, \bibinfo {author} {\bibfnamefont {W.}~\bibnamefont {Liu}}, \bibinfo {author} {\bibfnamefont {P.}~\bibnamefont {Yu}}, \bibinfo {author} {\bibfnamefont {H.}~\bibnamefont {Gao}}, \bibinfo {author} {\bibfnamefont {Z.}~\bibnamefont {Liu}}, \bibinfo {author} {\bibfnamefont {A.~M.}\ \bibnamefont {Rappe}}, \bibinfo {author} {\bibfnamefont {C.~L.}\ \bibnamefont {Kane}}, \bibinfo {author} {\bibfnamefont {E.~J.}\ \bibnamefont {Mele}}, \emph {et~al.},\ }\bibfield  {title} {\bibinfo {title} {Spatially dispersive circular photogalvanic effect in a {W}eyl semimetal},\ }\href@noop {} {\bibfield  {journal} {\bibinfo  {journal} {Nature materials}\ }\textbf {\bibinfo {volume} {18}},\ \bibinfo {pages} {955} (\bibinfo {year} {2019})}\BibitemShut {NoStop}%
\bibitem [{\citenamefont {Ma}\ \emph {et~al.}(2021)\citenamefont {Ma}, \citenamefont {Grushin},\ and\ \citenamefont {Burch}}]{ma2021topology}%
  \BibitemOpen
  \bibfield  {author} {\bibinfo {author} {\bibfnamefont {Q.}~\bibnamefont {Ma}}, \bibinfo {author} {\bibfnamefont {A.~G.}\ \bibnamefont {Grushin}},\ and\ \bibinfo {author} {\bibfnamefont {K.~S.}\ \bibnamefont {Burch}},\ }\bibfield  {title} {\bibinfo {title} {Topology and geometry under the nonlinear electromagnetic spotlight},\ }\href@noop {} {\bibfield  {journal} {\bibinfo  {journal} {Nature Materials}\ }\textbf {\bibinfo {volume} {20}},\ \bibinfo {pages} {1601} (\bibinfo {year} {2021})}\BibitemShut {NoStop}%
\bibitem [{\citenamefont {Flicker}\ \emph {et~al.}(2018)\citenamefont {Flicker}, \citenamefont {De~Juan}, \citenamefont {Bradlyn}, \citenamefont {Morimoto}, \citenamefont {Vergniory},\ and\ \citenamefont {Grushin}}]{flicker2018chiral}%
  \BibitemOpen
  \bibfield  {author} {\bibinfo {author} {\bibfnamefont {F.}~\bibnamefont {Flicker}}, \bibinfo {author} {\bibfnamefont {F.}~\bibnamefont {De~Juan}}, \bibinfo {author} {\bibfnamefont {B.}~\bibnamefont {Bradlyn}}, \bibinfo {author} {\bibfnamefont {T.}~\bibnamefont {Morimoto}}, \bibinfo {author} {\bibfnamefont {M.~G.}\ \bibnamefont {Vergniory}},\ and\ \bibinfo {author} {\bibfnamefont {A.~G.}\ \bibnamefont {Grushin}},\ }\bibfield  {title} {\bibinfo {title} {Chiral optical response of multifold fermions},\ }\href@noop {} {\bibfield  {journal} {\bibinfo  {journal} {Physical Review B}\ }\textbf {\bibinfo {volume} {98}},\ \bibinfo {pages} {155145} (\bibinfo {year} {2018})}\BibitemShut {NoStop}%
\bibitem [{\citenamefont {Zhang}\ \emph {et~al.}(2018{\natexlab{b}})\citenamefont {Zhang}, \citenamefont {Ishizuka}, \citenamefont {van~den Brink}, \citenamefont {Felser}, \citenamefont {Yan},\ and\ \citenamefont {Nagaosa}}]{zhang2018photogalvanic}%
  \BibitemOpen
  \bibfield  {author} {\bibinfo {author} {\bibfnamefont {Y.}~\bibnamefont {Zhang}}, \bibinfo {author} {\bibfnamefont {H.}~\bibnamefont {Ishizuka}}, \bibinfo {author} {\bibfnamefont {J.}~\bibnamefont {van~den Brink}}, \bibinfo {author} {\bibfnamefont {C.}~\bibnamefont {Felser}}, \bibinfo {author} {\bibfnamefont {B.}~\bibnamefont {Yan}},\ and\ \bibinfo {author} {\bibfnamefont {N.}~\bibnamefont {Nagaosa}},\ }\bibfield  {title} {\bibinfo {title} {Photogalvanic effect in {W}eyl semimetals from first principles},\ }\href@noop {} {\bibfield  {journal} {\bibinfo  {journal} {Physical Review B}\ }\textbf {\bibinfo {volume} {97}},\ \bibinfo {pages} {241118} (\bibinfo {year} {2018}{\natexlab{b}})}\BibitemShut {NoStop}%
\bibitem [{\citenamefont {Avdoshkin}\ \emph {et~al.}(2020)\citenamefont {Avdoshkin}, \citenamefont {Kozii},\ and\ \citenamefont {Moore}}]{avdoshkin2020interactions}%
  \BibitemOpen
  \bibfield  {author} {\bibinfo {author} {\bibfnamefont {A.}~\bibnamefont {Avdoshkin}}, \bibinfo {author} {\bibfnamefont {V.}~\bibnamefont {Kozii}},\ and\ \bibinfo {author} {\bibfnamefont {J.~E.}\ \bibnamefont {Moore}},\ }\bibfield  {title} {\bibinfo {title} {Interactions remove the quantization of the chiral photocurrent at {W}eyl points},\ }\href@noop {} {\bibfield  {journal} {\bibinfo  {journal} {Physical review letters}\ }\textbf {\bibinfo {volume} {124}},\ \bibinfo {pages} {196603} (\bibinfo {year} {2020})}\BibitemShut {NoStop}%
\bibitem [{\citenamefont {Wu}\ \emph {et~al.}(2024)\citenamefont {Wu}, \citenamefont {Guerci}, \citenamefont {Fu}, \citenamefont {Wilson},\ and\ \citenamefont {Pixley}}]{wu2024absence}%
  \BibitemOpen
  \bibfield  {author} {\bibinfo {author} {\bibfnamefont {A.-K.}\ \bibnamefont {Wu}}, \bibinfo {author} {\bibfnamefont {D.}~\bibnamefont {Guerci}}, \bibinfo {author} {\bibfnamefont {Y.}~\bibnamefont {Fu}}, \bibinfo {author} {\bibfnamefont {J.~H.}\ \bibnamefont {Wilson}},\ and\ \bibinfo {author} {\bibfnamefont {J.}~\bibnamefont {Pixley}},\ }\bibfield  {title} {\bibinfo {title} {Absence of quantization in the circular photogalvanic effect in disordered chiral {W}eyl semimetals},\ }\href@noop {} {\bibfield  {journal} {\bibinfo  {journal} {Physical Review B}\ }\textbf {\bibinfo {volume} {110}},\ \bibinfo {pages} {014201} (\bibinfo {year} {2024})}\BibitemShut {NoStop}%
\bibitem [{\citenamefont {Yu}\ \emph {et~al.}(2019)\citenamefont {Yu}, \citenamefont {Zang},\ and\ \citenamefont {Liu}}]{yu2019magnetic}%
  \BibitemOpen
  \bibfield  {author} {\bibinfo {author} {\bibfnamefont {J.}~\bibnamefont {Yu}}, \bibinfo {author} {\bibfnamefont {J.}~\bibnamefont {Zang}},\ and\ \bibinfo {author} {\bibfnamefont {C.-X.}\ \bibnamefont {Liu}},\ }\bibfield  {title} {\bibinfo {title} {Magnetic resonance induced pseudoelectric field and giant current response in axion insulators},\ }\href@noop {} {\bibfield  {journal} {\bibinfo  {journal} {Physical Review B}\ }\textbf {\bibinfo {volume} {100}},\ \bibinfo {pages} {075303} (\bibinfo {year} {2019})}\BibitemShut {NoStop}%
\bibitem [{\citenamefont {Yu}\ and\ \citenamefont {Liu}(2021)}]{yu2021pseudo}%
  \BibitemOpen
  \bibfield  {author} {\bibinfo {author} {\bibfnamefont {J.}~\bibnamefont {Yu}}\ and\ \bibinfo {author} {\bibfnamefont {C.-X.}\ \bibnamefont {Liu}},\ }\bibfield  {title} {\bibinfo {title} {Pseudo-gauge fields in dirac and weyl materials},\ }in\ \href@noop {} {\emph {\bibinfo {booktitle} {Semiconductors and Semimetals}}},\ Vol.\ \bibinfo {volume} {108}\ (\bibinfo  {publisher} {Elsevier},\ \bibinfo {year} {2021})\ pp.\ \bibinfo {pages} {195--224}\BibitemShut {NoStop}%
\bibitem [{\citenamefont {Ilan}\ \emph {et~al.}(2020)\citenamefont {Ilan}, \citenamefont {Grushin},\ and\ \citenamefont {Pikulin}}]{ilan2020pseudo}%
  \BibitemOpen
  \bibfield  {author} {\bibinfo {author} {\bibfnamefont {R.}~\bibnamefont {Ilan}}, \bibinfo {author} {\bibfnamefont {A.~G.}\ \bibnamefont {Grushin}},\ and\ \bibinfo {author} {\bibfnamefont {D.~I.}\ \bibnamefont {Pikulin}},\ }\bibfield  {title} {\bibinfo {title} {Pseudo-electromagnetic fields in 3d topological semimetals},\ }\href@noop {} {\bibfield  {journal} {\bibinfo  {journal} {Nature Reviews Physics}\ }\textbf {\bibinfo {volume} {2}},\ \bibinfo {pages} {29} (\bibinfo {year} {2020})}\BibitemShut {NoStop}%
\bibitem [{\citenamefont {Liu}\ \emph {et~al.}(2013)\citenamefont {Liu}, \citenamefont {Ye},\ and\ \citenamefont {Qi}}]{liu2013chiral}%
  \BibitemOpen
  \bibfield  {author} {\bibinfo {author} {\bibfnamefont {C.-X.}\ \bibnamefont {Liu}}, \bibinfo {author} {\bibfnamefont {P.}~\bibnamefont {Ye}},\ and\ \bibinfo {author} {\bibfnamefont {X.-L.}\ \bibnamefont {Qi}},\ }\bibfield  {title} {\bibinfo {title} {Chiral gauge field and axial anomaly in a {W}eyl semimetal},\ }\href@noop {} {\bibfield  {journal} {\bibinfo  {journal} {Physical Review B}\ }\textbf {\bibinfo {volume} {87}},\ \bibinfo {pages} {235306} (\bibinfo {year} {2013})}\BibitemShut {NoStop}%
\bibitem [{\citenamefont {Armitage}\ \emph {et~al.}(2018)\citenamefont {Armitage}, \citenamefont {Mele},\ and\ \citenamefont {Vishwanath}}]{armitage2018weyl}%
  \BibitemOpen
  \bibfield  {author} {\bibinfo {author} {\bibfnamefont {N.}~\bibnamefont {Armitage}}, \bibinfo {author} {\bibfnamefont {E.}~\bibnamefont {Mele}},\ and\ \bibinfo {author} {\bibfnamefont {A.}~\bibnamefont {Vishwanath}},\ }\bibfield  {title} {\bibinfo {title} {Weyl and dirac semimetals in three-dimensional solids},\ }\href@noop {} {\bibfield  {journal} {\bibinfo  {journal} {Reviews of Modern Physics}\ }\textbf {\bibinfo {volume} {90}},\ \bibinfo {pages} {015001} (\bibinfo {year} {2018})}\BibitemShut {NoStop}%
\bibitem [{\citenamefont {Kittel}\ and\ \citenamefont {McEuen}(1976)}]{kittel1976introduction}%
  \BibitemOpen
  \bibfield  {author} {\bibinfo {author} {\bibfnamefont {C.}~\bibnamefont {Kittel}}\ and\ \bibinfo {author} {\bibfnamefont {P.}~\bibnamefont {McEuen}},\ }\href@noop {} {\emph {\bibinfo {title} {Introduction to solid state physics}}},\ Vol.~\bibinfo {volume} {8}\ (\bibinfo  {publisher} {Wiley New York},\ \bibinfo {year} {1976})\BibitemShut {NoStop}%
\bibitem [{\citenamefont {Aharoni}(2000)}]{aharoni2000introduction}%
  \BibitemOpen
  \bibfield  {author} {\bibinfo {author} {\bibfnamefont {A.}~\bibnamefont {Aharoni}},\ }\href@noop {} {\emph {\bibinfo {title} {Introduction to the Theory of Ferromagnetism}}},\ Vol.\ \bibinfo {volume} {109}\ (\bibinfo  {publisher} {Clarendon Press},\ \bibinfo {year} {2000})\BibitemShut {NoStop}%
\bibitem [{\citenamefont {Aversa}\ and\ \citenamefont {Sipe}(1995)}]{aversa1995nonlinear}%
  \BibitemOpen
  \bibfield  {author} {\bibinfo {author} {\bibfnamefont {C.}~\bibnamefont {Aversa}}\ and\ \bibinfo {author} {\bibfnamefont {J.~E.}\ \bibnamefont {Sipe}},\ }\bibfield  {title} {\bibinfo {title} {Nonlinear optical susceptibilities of semiconductors: Results with a length-gauge analysis},\ }\href@noop {} {\bibfield  {journal} {\bibinfo  {journal} {Physical Review B}\ }\textbf {\bibinfo {volume} {52}},\ \bibinfo {pages} {14636} (\bibinfo {year} {1995})}\BibitemShut {NoStop}%
\bibitem [{\citenamefont {Ventura}\ \emph {et~al.}(2017)\citenamefont {Ventura}, \citenamefont {Passos}, \citenamefont {Dos~Santos}, \citenamefont {Lopes},\ and\ \citenamefont {Peres}}]{ventura2017gauge}%
  \BibitemOpen
  \bibfield  {author} {\bibinfo {author} {\bibfnamefont {G.}~\bibnamefont {Ventura}}, \bibinfo {author} {\bibfnamefont {D.}~\bibnamefont {Passos}}, \bibinfo {author} {\bibfnamefont {J.~L.}\ \bibnamefont {Dos~Santos}}, \bibinfo {author} {\bibfnamefont {J.~V.~P.}\ \bibnamefont {Lopes}},\ and\ \bibinfo {author} {\bibfnamefont {N.~M.}\ \bibnamefont {Peres}},\ }\bibfield  {title} {\bibinfo {title} {Gauge covariances and nonlinear optical responses},\ }\href@noop {} {\bibfield  {journal} {\bibinfo  {journal} {Physical Review B}\ }\textbf {\bibinfo {volume} {96}},\ \bibinfo {pages} {035431} (\bibinfo {year} {2017})}\BibitemShut {NoStop}%
\bibitem [{\citenamefont {Sipe}\ and\ \citenamefont {Shkrebtii}(2000)}]{sipe2000second}%
  \BibitemOpen
  \bibfield  {author} {\bibinfo {author} {\bibfnamefont {J.}~\bibnamefont {Sipe}}\ and\ \bibinfo {author} {\bibfnamefont {A.}~\bibnamefont {Shkrebtii}},\ }\bibfield  {title} {\bibinfo {title} {Second-order optical response in semiconductors},\ }\href@noop {} {\bibfield  {journal} {\bibinfo  {journal} {Physical Review B}\ }\textbf {\bibinfo {volume} {61}},\ \bibinfo {pages} {5337} (\bibinfo {year} {2000})}\BibitemShut {NoStop}%
\bibitem [{\citenamefont {Chen}\ \emph {et~al.}(2022)\citenamefont {Chen}, \citenamefont {Ye}, \citenamefont {Zou}, \citenamefont {Gu}, \citenamefont {Xu},\ and\ \citenamefont {Duan}}]{chen2022basic}%
  \BibitemOpen
  \bibfield  {author} {\bibinfo {author} {\bibfnamefont {H.}~\bibnamefont {Chen}}, \bibinfo {author} {\bibfnamefont {M.}~\bibnamefont {Ye}}, \bibinfo {author} {\bibfnamefont {N.}~\bibnamefont {Zou}}, \bibinfo {author} {\bibfnamefont {B.-L.}\ \bibnamefont {Gu}}, \bibinfo {author} {\bibfnamefont {Y.}~\bibnamefont {Xu}},\ and\ \bibinfo {author} {\bibfnamefont {W.}~\bibnamefont {Duan}},\ }\bibfield  {title} {\bibinfo {title} {Basic formulation and first-principles implementation of nonlinear magneto-optical effects},\ }\href@noop {} {\bibfield  {journal} {\bibinfo  {journal} {Physical Review B}\ }\textbf {\bibinfo {volume} {105}},\ \bibinfo {pages} {075123} (\bibinfo {year} {2022})}\BibitemShut {NoStop}%
\bibitem [{\citenamefont {Parker}\ \emph {et~al.}(2019)\citenamefont {Parker}, \citenamefont {Morimoto}, \citenamefont {Orenstein},\ and\ \citenamefont {Moore}}]{parker2019diagrammatic}%
  \BibitemOpen
  \bibfield  {author} {\bibinfo {author} {\bibfnamefont {D.~E.}\ \bibnamefont {Parker}}, \bibinfo {author} {\bibfnamefont {T.}~\bibnamefont {Morimoto}}, \bibinfo {author} {\bibfnamefont {J.}~\bibnamefont {Orenstein}},\ and\ \bibinfo {author} {\bibfnamefont {J.~E.}\ \bibnamefont {Moore}},\ }\bibfield  {title} {\bibinfo {title} {Diagrammatic approach to nonlinear optical response with application to {W}eyl semimetals},\ }\href@noop {} {\bibfield  {journal} {\bibinfo  {journal} {Physical Review B}\ }\textbf {\bibinfo {volume} {99}},\ \bibinfo {pages} {045121} (\bibinfo {year} {2019})}\BibitemShut {NoStop}%
\bibitem [{\citenamefont {Passos}\ \emph {et~al.}(2018)\citenamefont {Passos}, \citenamefont {Ventura}, \citenamefont {Lopes}, \citenamefont {Dos~Santos},\ and\ \citenamefont {Peres}}]{passos2018nonlinear}%
  \BibitemOpen
  \bibfield  {author} {\bibinfo {author} {\bibfnamefont {D.}~\bibnamefont {Passos}}, \bibinfo {author} {\bibfnamefont {G.}~\bibnamefont {Ventura}}, \bibinfo {author} {\bibfnamefont {J.~V.~P.}\ \bibnamefont {Lopes}}, \bibinfo {author} {\bibfnamefont {J.~L.}\ \bibnamefont {Dos~Santos}},\ and\ \bibinfo {author} {\bibfnamefont {N.}~\bibnamefont {Peres}},\ }\bibfield  {title} {\bibinfo {title} {Nonlinear optical responses of crystalline systems: Results from a velocity gauge analysis},\ }\href@noop {} {\bibfield  {journal} {\bibinfo  {journal} {Physical Review B}\ }\textbf {\bibinfo {volume} {97}},\ \bibinfo {pages} {235446} (\bibinfo {year} {2018})}\BibitemShut {NoStop}%
\end{thebibliography}%

\end{document}